\documentclass[aps,pra,twocolumn,superscriptaddress]{revtex4-2}
\usepackage{graphicx}
\usepackage[hidelinks]{hyperref}
\usepackage{bm}
\usepackage{multirow}
\usepackage{mathtools, amsthm, amsfonts, amssymb}
\usepackage{soul}
\newcommand{\e}{\operatorname{e}}

\begin{document}


\title{A mass invariant in a compressible turbulent medium}

\author{Pierre Dumond}
\email[]{pierre.dumond@ens-lyon.fr}
\affiliation{CRAL, Ecole Normale Supérieure de Lyon, Université de Lyon, UMR CNRS 5574, F-69364 Lyon Cedex 07, France}

\author{Jérémy Fensch}
\affiliation{CRAL, Ecole Normale Supérieure de Lyon, Université de Lyon, UMR CNRS 5574, F-69364 Lyon Cedex 07, France}

\author{Gilles Chabrier}
\affiliation{CRAL, Ecole Normale Supérieure de Lyon, Université de Lyon, UMR CNRS 5574, F-69364 Lyon Cedex 07, France}
\affiliation{School of Physics, University of Exeter, Exeter, EX4 4QL, UK}

\author{Etienne Jaupart}
\affiliation{CEA-DAM-DIF, F-91297 Arpajon, France}

\date{\today}

\begin{abstract}
    Predicting the measurable statistical properties of density fluctuations in a supersonic compressible turbulent flow is a major challenge in physics. In 1951, Chandrasekhar derived an invariant under the assumption of the statistical homogeneity and isotropy of the turbulent density field and stationarity of the background density. Recently, Jaupart \& Chabrier (2021) extended this invariant to non-isotropic flows in a time-evolving background and showed that it has the dimension of a mass. This invariant \(M_{\rm inv}\) is defined by \(M_{\rm inv} = \mathbb{E}(\rho)\text{Var}\left(\frac{\rho}{\mathbb{E}(\rho)}\right)(l_{\rm c}^\rho)^3\) where \(\rho\) is the density field and \(l_{\rm c}^\rho\) is the correlation length. In this article, we perform numerical simulations of homogeneous and isotropic compressible turbulence to test the validity of this invariant in a medium subject to isotropic decaying turbulence. We study several input configurations, namely different Mach numbers, injection lengths of turbulence and equations of state. We confirm that \(M_{\rm inv}\) remains constant during the decaying phase of  turbulence. Furthermore, we develop a theoretical model of the density field statistics which predicts without any free parameter the evolution of the correlation length with the  variance of the logdensity field beyond the Gaussian random field assumption for the logdensity. Noting that \(M_{\rm inv}\) is independent of the Mach number, we show that this invariant can be used to relate the non-Gaussian evolution of the logdensity probability distribution function to its variance with no free parameters.
\end{abstract}


\maketitle

\section{Introduction}

Properly understanding the statistical properties of compressible turbulence remains a major problem in physics. Most attempts have been limited to the study of weakly compressible turbulence, i.e. 3D Mach number \(\mathcal{M}=\sqrt{\langle v^2\rangle}/c_{\rm s}\leq 1\) where \(\sqrt{\langle v^2\rangle}\) is the 3D velocity dispersion and \(c_{\rm s}\) is the sound speed, both on the theoretical side (\cite{Bayly_DensityVariationsWeakly1992,Bertoglio_TwopointClosuresWeakly2001}, and \cite{Zhou_TurbulenceTheoriesStatistical2021} for a review) and on the numerical side (e.g. \cite{Wang_StatisticsStructuresPressure2013, Wang_KineticEnergyTransfer2018}). Little attention has been paid to supersonic compressible turbulence, mainly because in this case none of the usual closure equations (e.g. Eddy Damped Quasi-Normal Markovian or, Direct Interaction Approximation) can be used to derive the statistical properties of interest for the flow. Therefore, our current understanding of compressible turbulence is based essentially on phenomenology, notably the description of the field of density fluctuations as being the result of the interaction of a large number of random shocks, allowing the use of the Central Limit Theorem for multiplicative independent random processes \citep{Vazquez-Semadeni_HierarchicalStructureNearly1994a,Squire_DistributionDensitySupersonic2017}. However, there is no exact result to describe the commonly measured statistics of the flow (probability density function [PDF], power spectra). Some exact relations do exist \citep{Galtier_ExactRelationCorrelation2011}, but it is difficult to extract from them a prediction of the statistical evolution of the flow properties.

Assuming ergodicity, stationarity of the background density and statistical homogeneity\footnote{We recall that a stochastic field is statistically homogeneous if all its statistical properties are invariant under space translations. Not to be confused with spatial homogeneity.} and isotropy of the turbulent flow, Ref. \cite{Chandrasekhar_FluctuationsDensityIsotropic1951} derived a temporal invariant based on the continuity equation. He also had to assume that the cross-correlation \(C_{\rho, \rho\bm{v}}(|\bm{q}|)\) between the density \(\rho\) and the momentum \(\rho\bm{v}\) decays faster than \(1/|\bm{q}|^2\) at infinity, where \(\bm{q}\) denotes the spatial distance.
Ref. \cite{Jaupart_GeneralizedTransportEquation2021a} has recently extended this invariant to non-isotropic turbulence and evolving statistical mean density of the flow \(\mathbb{E}(\rho)\), yielding the following form:
\begin{equation}
    \label{eq_M_inv_def}
    M_{\rm inv} = \mathbb{E}(\rho)(t)\text{Var}\left(\frac{\rho}{\mathbb{E}(\rho)}\right)_t(l_{\rm c}^\rho)_t^3 = \text{const},
\end{equation}
where \(l_{\rm c}^\rho\) is the correlation length, defined as:
\begin{equation}
    l_{\rm c}^\rho = \left(\frac{1}{8C_\rho(0)}\int_{\mathbb{R}^3} C_\rho(\bm{q}) {\rm d}^3\bm{q}\right)^{1/3},
    \label{eq:lc}
\end{equation}
where \(C_\rho\) is the autocovariance function (ACF) of the density field. We recall that, by definition, \(C_\rho(0)\) is the variance of the density field. Note that the basic mathematical expressions of the invariant derived by \cite{Chandrasekhar_FluctuationsDensityIsotropic1951} and by \cite{Jaupart_GeneralizedTransportEquation2021a} are the same. The only difference between the two is the context within which they are applied. The invariant derived by Ref. \cite{Chandrasekhar_FluctuationsDensityIsotropic1951} applies to homegeneous isotropic turbulence with constant background average density, while the one derived by Ref. \cite{Jaupart_GeneralizedTransportEquation2021a} applies to homogeneous non-isotropic background with time-evolving background average density.

In this paper we mainly verify the invariant in the context studied by \cite{Chandrasekhar_FluctuationsDensityIsotropic1951}, i.e. for isotropic turbulence with no time variation of the statistical mean density of the flow and no gravity. We will briefly consider the case of a time-varying background density in \S III.C and leave the study of the invariant in anisotropic turbulent flows and in the presence of gravity for a future study.

In Sec. \ref{Sec_Simulations}, we first present the setup we use to numerically verify this invariant, making sure that the hypotheses it relies on are verified in the simulated turbulent flow. We show in Sec. \ref{Sec_Results} that this quantity is indeed constant over time in decaying turbulence. In Sec. \ref{Sec_Model} we derive an analytical model that describes well the statistical properties of the density field and allows to predict the variation of the correlation length with the Mach number without any fitting parameters. Finally, in Sec. \ref{Sec_Intermittency}, we show that this invariant can be used to determine analytically the evolution with the Mach number of the parameter \(T\) introduced by Ref. \cite{Hopkins_ModelNonlognormalDensity2013a} to describe the non-Gaussianity of the logdensity PDF.

\section{Simulations}
\label{Sec_Simulations}

\subsection{Numerical setup}

We use the hydrodynamical code RAMSES \citep{Teyssier_CosmologicalHydrodynamicsAdaptive2002} without adaptive mesh refinement (fixed cartersian grid). The boundary conditions are periodic. The numerical method is based on a second-order Godunov solver scheme. The solver is the Harten–Lax–van Leer-Contact (HLLC) approximate Riemann solver \citep{Toro_RiemannSolversNumerical2009}. The initial conditions consist of a fluid of atomic hydrogen at rest. The fluid follows an isothermal or polytropic equation of state. The turbulence is forced using the Ornstein-Uhlenbeck
forcing on the acceleration \citep{Eswaran_ExaminationForcingDirect1988a,Schmidt_NumericalDissipationBottleneck2006a,Schmidt_NumericalSimulationsCompressively2009}. The equations of conservation of mass and momentum that are solved are the following:
\begin{align}
    \frac{\partial \rho}{\partial t}+\nabla \cdot(\rho \bm{v}) & =0, \\
    \rho\left(\frac{\partial \bm{v}}{\partial t}+(\bm{v} \cdot \bm{\nabla}) \bm{v}\right) & =-c_{\rm s}^2\bm{\nabla}\rho + \rho \bm{f}.
\end{align}
The viscosity is not modelled explicitly, but acts implicitly on the flow through the numerical grid viscosity. The energy equation is not solved explicitly. It is replaced by a simple polytropic equation of state \(P\propto \rho^\gamma\) with \(\gamma\) the polytropic index. In this study, we will consider two cases, \(\gamma=1\) (isothermal process) and  \(\gamma=5/3\) (isentropic process for monoatomic gas).
The Fourier modes $\hat{\bm{f}}(\bm{k}, t)$ of the turbulence driving acceleration field \(\bm{f}\) follow the following stochastic differential equation:
\begin{equation}
\mathrm{d} \hat{\bm{f}}(\bm{k}, t)=-\hat{\bm{f}}(\bm{k}, t) \frac{\mathrm{d} t}{T_{\rm OU}}+F_0(\bm{k}) \bm{P}^\zeta(\bm{k}) \mathrm{d} \bm{W}_t .
\end{equation}
In this equation, $\mathrm{d} t$ is the integration time step and $T_{\rm OU}$ is the autocorrelation timescale. As usually done in such numerical simulations \citep[e.g.][]{Schmidt_NumericalDissipationBottleneck2006a}, we set \(T_{\rm OU}\) to the turbulent crossing time \(T_{\rm cross}=L_{\rm inj}/\sqrt{\langle v^2\rangle}\), where \(L_{\rm inj}\) is the injection length of turbulence and \(\sqrt{\langle v^2\rangle}\) the turbulent velocity dispersion.
The weighting function of the driving modes $F_0$ allows the turbulence to be driven only within a precise range of spatial scales. In our work, we inject the turbulence isotropically in most runs (unless stated otherwise) between \(L_{\rm box}/6\) and \(L_{\rm box}/8\):
\begin{equation}
    \label{eq_forcing_scale}
F_0(k)=\left\{\begin{array}{l}
1-\left(\frac{L_{\rm box}|\bm{k}|}{2 \pi}-7\right)^2 \text { if } 6<\frac{L_{\rm box}|\bm{k}|}{2 \pi}<8 \\
0 \text { if not. }
\end{array}\right.
\end{equation}
In most of the turbulent simulations in the literature, turbulence is injected at larger scale, typically \(L_{\rm box}/2\) (see discussion \ref{Sec_Subboxes}). Here, we chose to inject at smaller scale to ensure that the size of the simulation box is large enough compared to the correlation length (see Sec. \ref{Sec_Subboxes}). Assuming ergodicity, this ensure that spatial averages are good estimate of the statistical average (see discussion in Sec. \ref{Sec_Results}).

The projection operator $\bm{P}^\zeta$ is a weighted sum of the components of the Helmholtz decomposition of compressive versus solenoidal modes \citep{Federrath_ComparingStatisticsInterstellar2010}:
\begin{align}
    P_{i j}^\zeta(k) &=\zeta P_{i j}^{\perp}(k)+(1-\zeta) P_{i j}^{\|}(k) \\
    &=\zeta \delta_{i j}+(1-2 \zeta) \frac{k_i k_j}{|k|^2}, \nonumber
\end{align}
where $\delta_{i j}$ is the Kronecker symbol, and $P_{i j}^{\perp}=\delta_{i j}-k_i k_j / k^2$ and $P_{i j}^{\|}=k_i k_j / k^2$ are the fully solenoidal and fully compressive projection operators, respectively. The projection operator is used to construct a purely solenoidal force field by setting the solenoidal fraction $\zeta=1$, which is used to drive turbulence in an incompressible system \citep{Eswaran_ExaminationForcingDirect1988a}. In the following we have chosen \(\zeta=0.5\), which corresponds to the energy equipartition of the velocity field: 1/3 compressive and 2/3 solenoidal.

\begin{figure}
    \centering
    \includegraphics[width=\columnwidth]{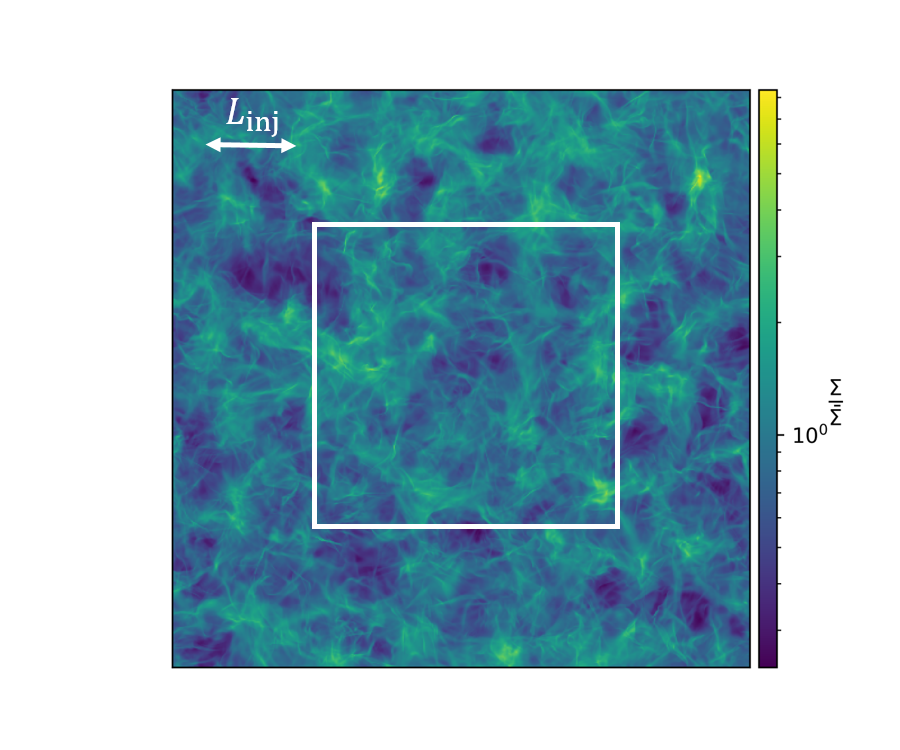}
    \caption{Projected density \(\Sigma\) of the simulation with the turbulence driven at Mach 3.5 and injected at \(L_{\rm box}/7\). The region within the white square is the one in which the study of the invariant is preformed in order to avoid the effect of the boundary conditions (see Sec. \ref{Sec_Subboxes}). The horizontal double arrow shows the size of the injection scale \(L_{\rm inj}\).}
    \label{fig_proj_density}
 \end{figure}

We carried out simulations at different large scale Mach numbers: \(\mathcal{M}=2.5\), 3.5, 5, 7 and 10 to test the invariant under different conditions. The projected density of the simulation at \(\mathcal{M}=3.5\) is shown in Fig. \ref{fig_proj_density}. We also consider several spatial resolutions and different injection lengths to test the convergence of the results. The list of the various simulations is given in Table \ref{tab_simus_list}. We also give the measured $\sigma_s^2$, which is a more primitive quantity of interest, and which depends on Mach and the type of forcing and equation of state.

\begin{table}
\begin{center}
\begin{tabular}{cccccc}
    \(\mathcal{M}\) & $\gamma$ & \(L_{\rm inj}\) &  \(N_{\rm cells}\) & $\sigma_s^2$ & Fiducial\\
    \hline \hline \multirow{5}{*}{ 2.5 } & \multirow{5}{*}{ 1 } & \multirow{3}{*}{ \(L_{\rm box}/7\)} &  \(1024^3\) & 1.37 & \\
    & & & \(2048^3\)  & 1.34 & \\\cline{3-6}
    & & \multirow{2}{*}{ \(L_{\rm box}/14\)}  &\(1024^3\) & 1.37 & \\
    & &  &\(2048^3\) & 1.36 & \\
    \hline \multirow{4}{*}{ 3.5 }  & \multirow{4}{*}{ 1 }& \multirow{3}{*}{ \(L_{\rm box}/7\)} & \(512^3\) & 2.17 & \\
    & & & \(1024^3\) & 2.07 & \\
    & & & \(2048^3\) & 2.04 & * \\\cline{3-6}
    & & \(L_{\rm box}/14\)  &\(1024^3\) & 2.01 & \\
    \hline \multirow{3}{*}{5} & \multirow{3}{*}{1 } & \multirow{3}{*}{ \(L_{\rm box}/7\)} &  \(512^3\) & 3.23 & \\
    & & & \(1024^3\) & 2.94 & \\
    & & & \(2048^3\) & 2.83 & *\\
    \hline \multirow{5}{*}{ 7.5 } & \multirow{4}{*}{ 1} & \multirow{3}{*}{ \(L_{\rm box}/7\)} & \(512^3\) & 4.39 & \\
    & & & \(1024^3\) & 4.19 & \\
    & & & \(2048^3\) & 4.01 & *\\\cline{3-6}
    & & \(L_{\rm box}/2\) & \(512^3\) & 4.39 & \\ \cline{2-6}
    & 5/3 & \(L_{\rm box}/7\) & \(1024^3\) & 2.70 \\
    \hline \multirow{3}{*}{10} & \multirow{3}{*}{1 } & \multirow{3}{*}{ \(L_{\rm box}/7\)} &  \(512^3\) & 5.70 & \\
    & & & \(1024^3\) & 5.44 & \\
    & & & \(2048^3\) & 4.95 & *\\
\end{tabular}
\end{center}
\caption{List of simulations. From left to right: the 3D Mach number \(\mathcal{M}\), the  polytropic index \(\gamma\), the injection length \(L_{\rm inj}\), the number of cells \(N_{\rm cells}\) and the variance of the logdensity field \(\sigma_s^2\). The \(*\) denote the fiducial run used in this study.}
\label{tab_simus_list}
\end{table}

 To test the temporal invariance of the quantity \(M_{\rm inv}\) [see Eq. (\ref{eq_M_inv_def})] suggested by Ref. \cite{Chandrasekhar_FluctuationsDensityIsotropic1951} and Ref. \cite{Jaupart_GeneralizedTransportEquation2021a}, we first drive the turbulence for at least 2 { turbulent} crossing times \(T_{\rm cross}\) before stopping the driving. This ensures that the turbulence is fully developed when the driving  is stopped \citep{Federrath_UniversalitySupersonicTurbulence2013}. During the turbulence decay phase, we measure \(M_{\rm inv}\) and verify whether it remains constant or not.

 \subsection{Sub-boxes}
\label{Sec_Subboxes}

Since the boundary conditions of the simulation domain are periodic, the periodicity could create artificial large scale correlations which could prevent the decay of the correlation between \(\rho\) and \(\rho\bm{v}\), which is one of the main assumptions for the quantity \(M_{\rm inv}\) to be time invariant. Furthermore, it can be shown that the correlation length measured in a periodic box with a periodic estimator is 0 \citep{Jaupart_StatisticalPropertiesCorrelation2022}. This suggests that any estimate of the correlation length in a periodic box will be inaccurate.
To overcome these problems, we study only a subsystem of the whole simulation domain. The size of the sub-box should be (i) small enough compared with the whole box to be unaffected by the boundary conditions and (ii) large enough compared with the correlation length to resolve it properly. We choose a sub-box with a volume of 1/8 of the whole box. We verified that changing the volume from 1/8 to 1/64 of the whole box does not affect qualitatively the results. All further analysis will be carried out in such a system.

In this sub-box, while the volumic mean density $\bar \rho$ is not constant but exhibits small variations (of the order of 5-10\%), the statistical mean density \(\mathbb{E}(\rho)\) is constant. It is equal to the mass contained in the periodic box divided by its volume. Furthermore, the anisotropy introduced by the cubic geometry of the subbox does not affect the calculation of the correlation length. In fact, this anisotropy has an effect only at the boundary of the subbox and thus on scales much larger than the correlation length.
The invariant that is tested is thus the one derived by Ref. \cite{Chandrasekhar_FluctuationsDensityIsotropic1951} for which the statistical mean density $\mathbb{E}(\rho)(t)$ does not change with time and the flow is statistically isotropic, in contrast to the more general derivation of Ref. \cite{Jaupart_GeneralizedTransportEquation2021a}.
\subsection{Computation of the correlation length}

\begin{figure*}
    \centering
    \includegraphics[width=\textwidth]{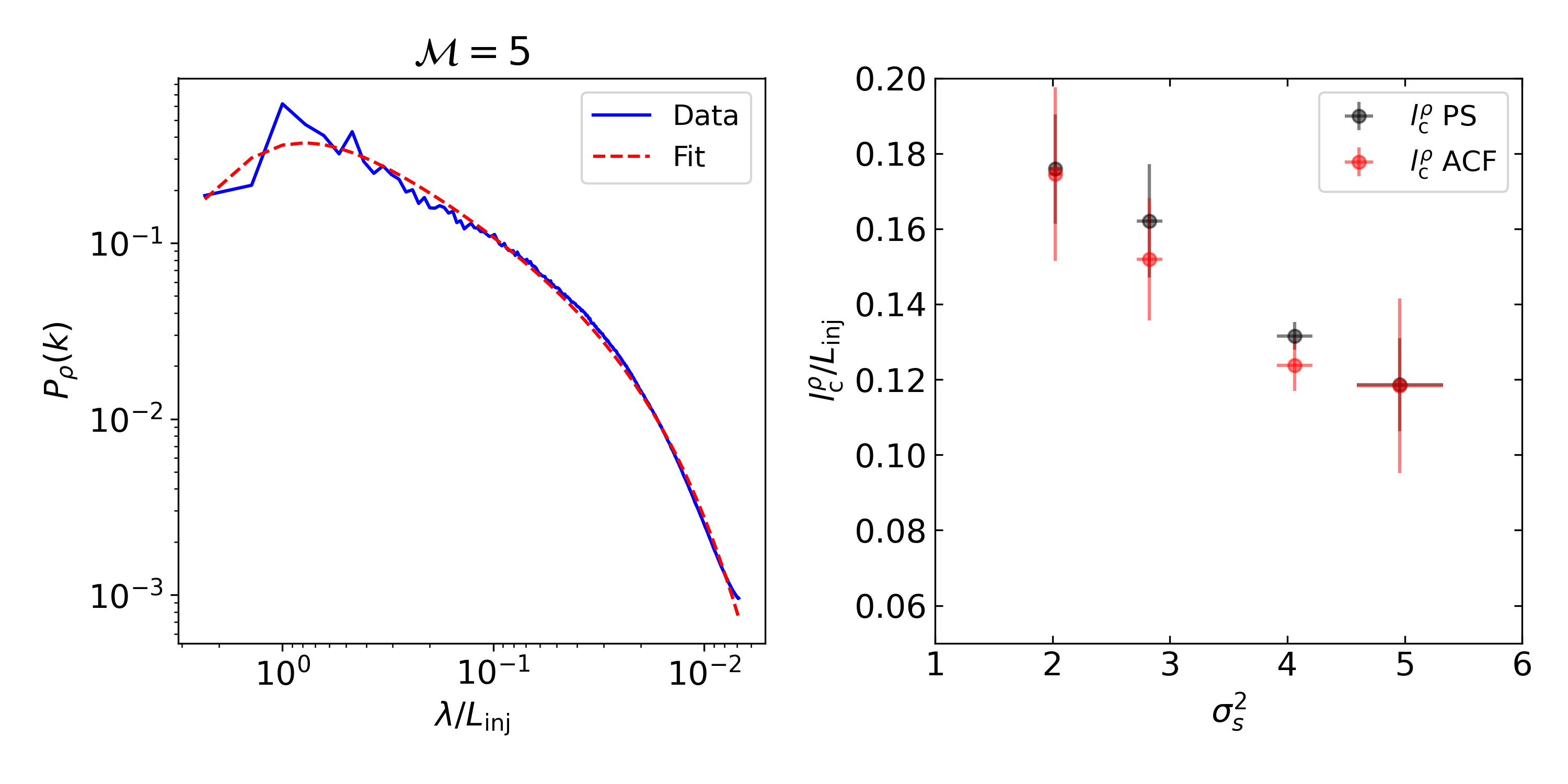}
    \caption{Left: 1D density Power spectrum (blue) and fit using Eq. \ref{eq_fit_PS_rho} as a function of the wavenumber \(\lambda=2\pi/k\) for one of the output in 2048\(^3\) simulation at Mach 5 during the stationary regime. Right: Correlation length measured in the 2048\(^3\) simulations by the integration ACF to its first zeros (red) and by the value at 0 of the power spectrum (black). The error bars correspond to the $\pm 1\sigma$ estimated from the time variations of the measured correlation length.}
    \label{fig_lc_methods}
 \end{figure*}

To verify the time invariance of \(M_{\rm inv}\), a major difficulty is to compute the correlation length of the density field. Under the assumption of ergodicity and statistical homogeneity, the ACF of a random field \(X\) can be calculated from its power spectrum \(|\mathcal{F}(X - \mathbb{E}(X))|^2\) using the Wiener-Khintchine theorem:
\begin{align}
    \label{eq_def_ACF}
    C_X(\bm{q}) &= \mathbb{E}(X(\bm{r})X(\bm{r}+\bm{q})) - \mathbb{E}(X)^2, \nonumber \\
    &= \mathcal{F}^{-1}(|\mathcal{F}(X - \mathbb{E}(X))|^2)(\bm{q}),
\end{align}
where \(\mathcal{F}\) is the Fourier transform operator and \(\mathbb{E}\) the statistical average. Numerically, however, the correlation function  exhibits numerical noise at large distances which prevents its integral over the volume [see Eq. (\ref{eq:lc})] from converging. These spurious oscillations are probably due to the limited statistics at large distance.

To avoid this problem, various estimate of the correlation length have been suggested \citep{Batchelor_TheoryHomogeneousTurbulence1953a, Bialy_DrivingScaleDensity2020,Jaupart_GeneralizedTransportEquation2021a,Jaupart_StatisticalPropertiesCorrelation2022}. In Appendix \ref{App_correlation_length}, we detail these possible estimates. However, none of them can be applied to the present case, either because they are too imprecise or because they need too many computational ressources.
Here, to compute the correlation length, we choose to stop the volumic integration of the ACF at its first zero, assuming that that the ACF at larger distance contains mostly noise. Due to the rapid decay of the ACF at large distance, the cutoff of ACF at its first zero does not impact too much the obtained value of the autocorrelation length.

To check that the value of the correlation length obtained by this method is accurate, we confront it to an other method based on the value at 0 of the power spectrum \citep{Jaupart_StatisticalPropertiesCorrelation2022}. However, we do not have access to this value because of the finite size of the box within which the measurements are done. To obtain this value, we thus have to impose a functional form of the power spectrum. We chose the following one with the slope of the inertial range \(2\eta\) and the dissipation scale \(\beta\):
\begin{align}
    \label{eq_fit_PS_rho}
    P_\rho^{\rm 1D}(k) &\propto k^2 P_\rho^{\rm 3D}(k) &\propto \frac{\e^{-k\beta}}{\left(1+(\frac{kL_{\rm inj}}{2\pi})^2\right)^\eta}k^2,
\end{align}

In Figure \ref{fig_lc_methods}, we compare the value of the correlation length computed from the fit of the power spectrum and the integration to the first zeros of the ACF. These computation are performed during the stationary regime of the 2048\(^3\) simulations. Both estimates are in good agreement with each other. A comparison between our measurement based on the integration to the first zeros of the ACF and on the ergodic theory is presented in Appendix \ref{App_correlation_length} for simulations at 1024\(^3\). We have not made this comparison for our 2048\(^3\) because of the high computational cost needed to use the estimate of the correlation length with the ergodic method.

A key condition to be sure that the correlation length is well resolved is that it must be significantly smaller than the size of the system under study. From our numerical tests, we have verified that the correlation length is well resolved when it is at most 1/15 of the size of the system within which the measures are done (see Sec. \ref{Sec_Subboxes}). In addition, we find that the correlation length must be resolved by at least 10 cells in order to be properly converged (see Appendix \ref{App_num_convergence}).

\section{Temporal invariance of \(M_{\rm inv}\)}
\label{Sec_Results}

\subsection{Steady state background}
\label{Sec_steady_state_back}

In this section we present the results of our numerical test of the time invariance of \(M_{\rm inv}\) in a steady state background for {two different equation of states. We first consider an isothermal equation of state (\(\gamma=1\)) before studying a polytropic one (\(\gamma=5/3\))}.

\begin{figure*}
    \centering
    \includegraphics[width=\textwidth]{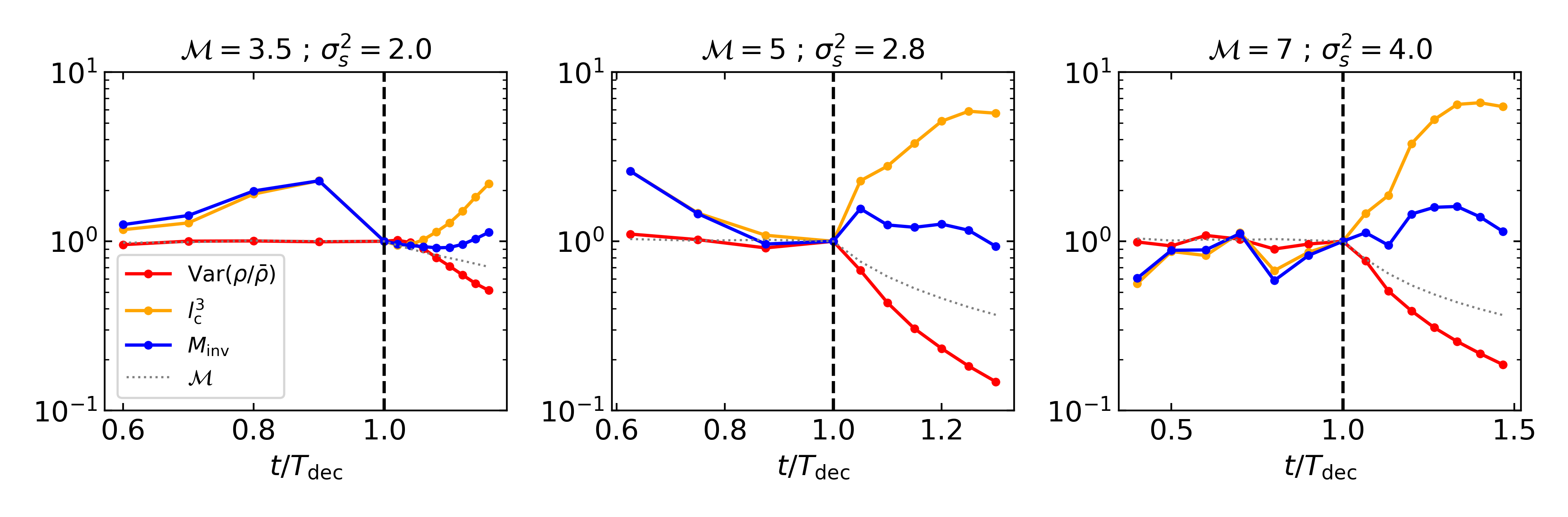}
    \caption{Evolution of mass \(M_{\rm inv}\) (blue), correlation length \(l_{\rm c}^3\) (orange), density variance \(\mathrm{Var}(\rho/\mathbb{E}(\rho))\) (red) and the Mach number (black dashed) for three Mach numbers in the 2048\(^3\) simulations. {The equation of state is isothermal in these three simualtions}. The black dotted line corresponds to the time \(T_{\rm dec}\) at which the turbulence forcing is switched off. The integration is stopped when the correlation length is no longer well resolved, and only the times after which the turbulence is fully developed are shown. This corresponds to \(t>2 T_{\rm cross}\) (see text). All the four quantities shown are normalized to their value at the time the turbulence forcing is switched off. The time is normalized to the time \(T_{\rm dec}\).}
    \label{fig_invariant_demo}
\end{figure*}

In Fig. \ref{fig_invariant_demo}, for the isothermal equation of state, we show that, during the turbulence decay phase, \(M_{\rm inv}\) remains relatively constant (less than 30\% variation) compared to the decrease of the variance and the increase of the cube of correlation length, which vary between factor 2 and 10.
We stop the integration of the simulation when the latter is no longer well resolved.

Fluctuations in the evaluations of the invariant reflect fluctuations in the estimates of statistical quantities. The biggest fluctuations come from the evaluations of the correlation length, which is notoriously delicate to estimate (see Sec. C and Ref. \cite{Papoulis_ProbabilityRandomVariables2002}) . These fluctuations are the result of averaging over a finite volume. This is equivalent to saying that we have access to a finite number of independent samples from which to estimate the various statistics. The greater the ratio \(l_{\rm c}/L_{\rm box}\), the greater the fluctuations of the estimates around the true statistical average.

Since the true statistical quantity \(M_{\rm inv}\) is known to be constant by definition during the stationary phase, the fluctuations of the estimate of \(M_{\rm inv}\) in Fig. \ref{fig_invariant_demo} reflect the typical fluctuations generated by our estimator around this constant average and can be estimated. They are of the order of a factor 2 to 3.
Since the fluctuations of the measured invariant during the decaying phase are smaller or of the order of those measured in the stationary phase, we can conclude that the true statistical quantity \(M_{\rm inv}\) is constant during the decaying phase.

\begin{figure*}
    \centering
    \includegraphics[scale=0.5]{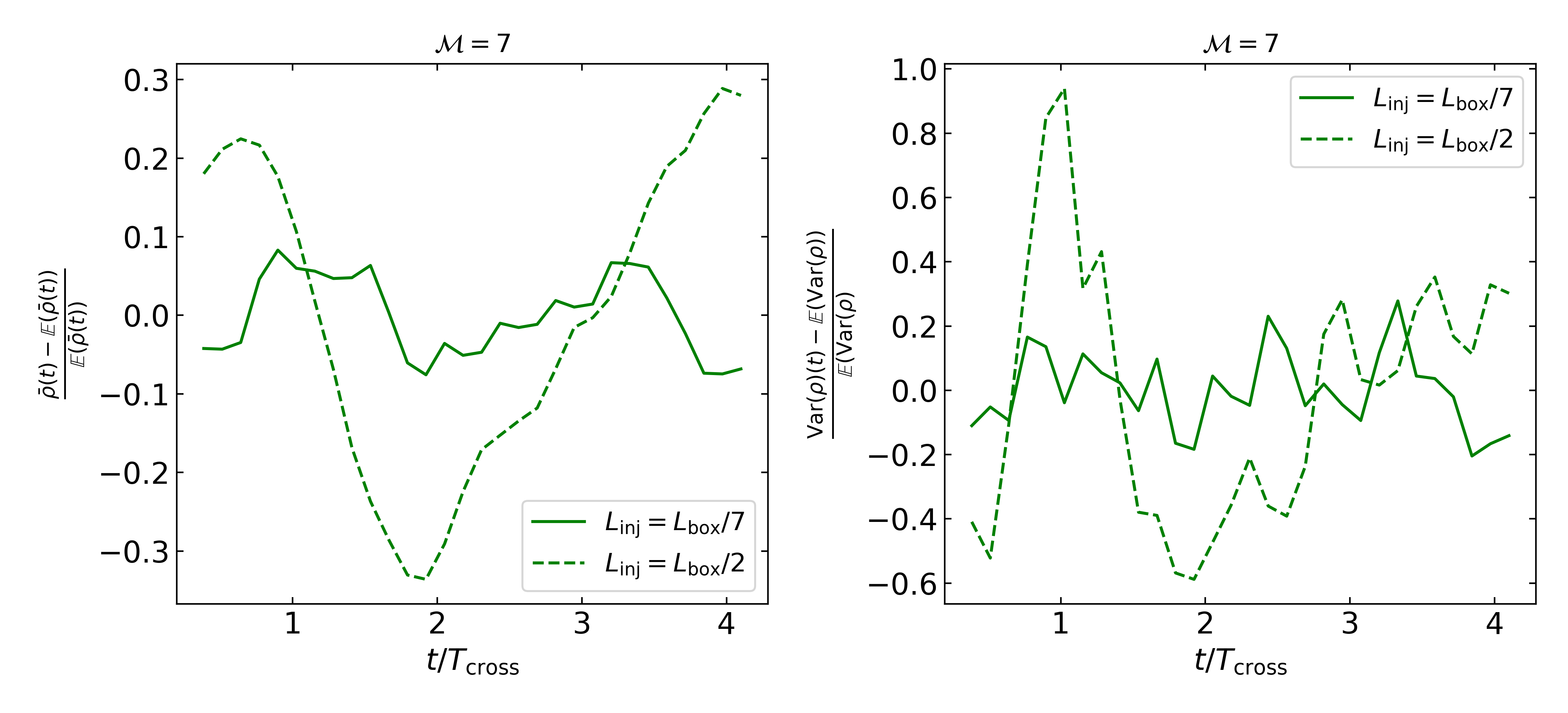}
    \caption{Evolution of the relative fluctuation of the mean density of the gas (left) and its variance (right) in the sub-box for a turbulence driving injected at \(L_{\rm inj}=L_{\rm box}/7\) (solid) and \(L_{\rm inj}=L_{\rm box}/2\) (dash) from the simulations at \(\mathcal{M}=7\) and 512\(^3\). }
    \label{fig_mean_density_fluctuations}
 \end{figure*}

In our numerical setup, the injection of turbulence at a scale smaller than the size of the box ensures that the volumic average over the measurement box of the quantities of interest is a good estimate of the true statistical value.
This is confirmed by the small fluctuations of the mean density and variance during the statistically stationary regime before the turbulence driving is switched off. The relative fluctuations of these quantities are of the order of 5\% - 10\%. We compared this with a simulation in which the turbulence is driven at \(L_{\rm box}/2\) and \(\mathcal{M}=7\). In that case, the relative fluctuations of the mean density and the density variance are of the order of 30\% (see Fig. \ref{fig_mean_density_fluctuations}), and the variance \(\sigma_\rho^2\) can vary by up to a factor of 2 during the simulation.

This demonstrates that turbulence injection at a scale significantly smaller than the box size is necessary to obtain good estimates of the statistical quantities of interest when performing volumic averaging. If the turbulence is injected at the box scale, an accurate estimate of the statistical mean can still be made by considering the time average over several turbulent crossing times. However, this method is limited to a statistically stationary turbulence regime.
To study the statistical properties of non-stationary turbulent flows, e.g. decaying or self-gravitating and thus fragmenting, it is therefore of paramount importance to have sufficient statistical realisations in volume, i.e. \(L_{\rm inj}\ll L_{\rm box}\), contrary to what is usually done in the literature where turbulence is usually injected at scales of the order of \(L_{\rm box}/2\). Their results may not be statistically accurate.

\begin{figure}
    \centering
    \includegraphics[width=0.8\columnwidth]{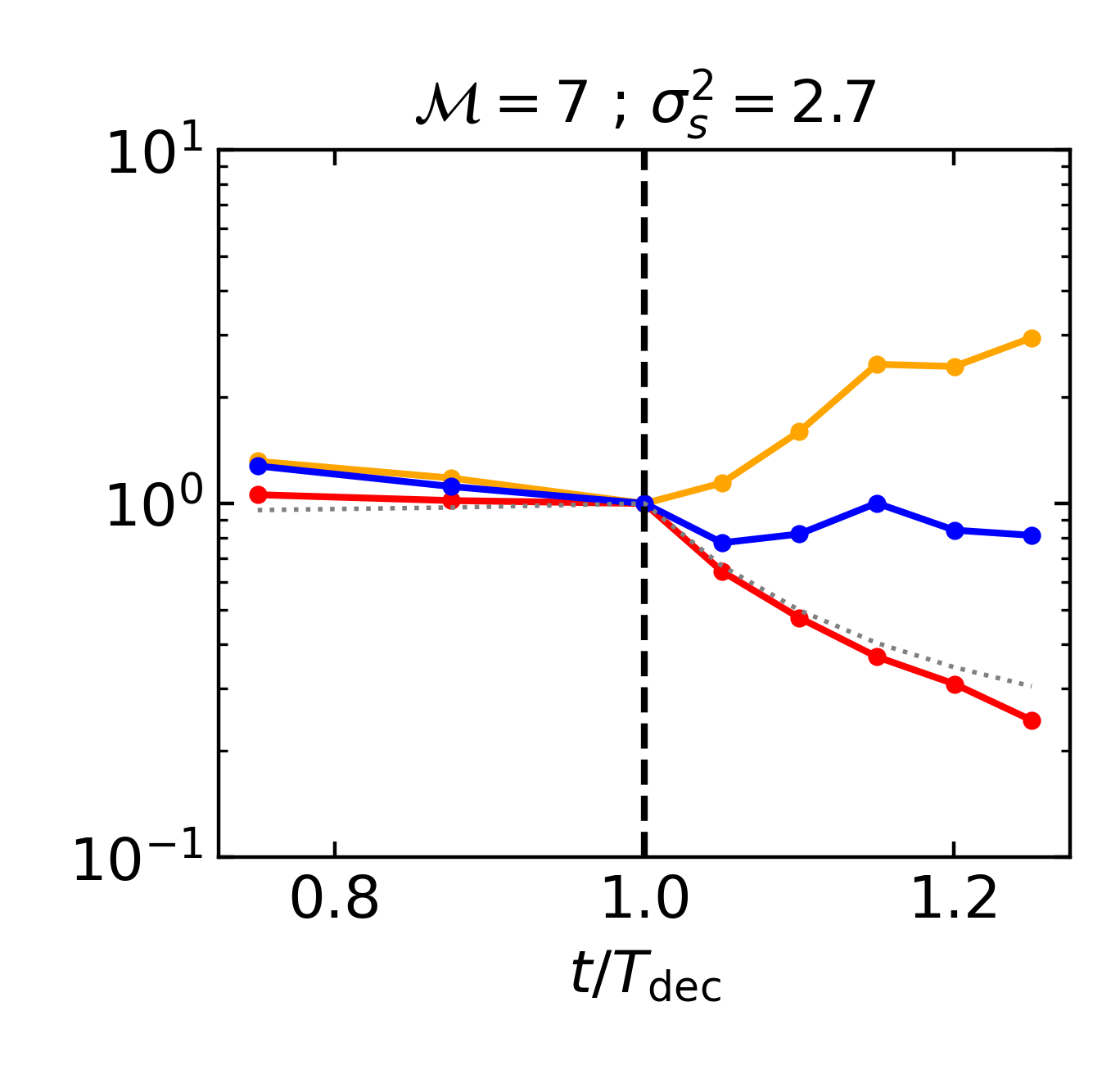}
    \caption{Same as Figure \ref{fig_invariant_demo} but for a polytropic equation of state with \(\gamma=5/3\). The simulation is run at \(\mathcal{M}=7\) with a 1024\(^3\) resolution. }
    \label{fig_invariant_demo_addiab}
 \end{figure}

Moreover, the invariant $M_{\rm inv}$ is a dynamics invariant, independent of the equation of state, as it is simply based on the conservation of mass equation. For the sake of completeness, and for reasons that will become apparent in Sec. \ref{Sec_inv_expansion}, we have also chosen to verify the invariance of $M_{\rm inv}$ using an polytropic equation of state with $\gamma =5/3$. It corresponds to the usual polytropic index for a monoatomic gas. In Fig. \ref{fig_invariant_demo_addiab}, and as in Fig. \ref{fig_invariant_demo}, we plot the evolution of the variance $\mathrm{Var}(\rho/\mathbb{E}(\rho))$, the cube of the correlation length \(l_{\rm c}^3\) and the invariant \(M_{\rm inv}\). During the decay, each term exhibits statistical fluctuations but the invariant $M_{\rm inv}$ remains constant with a standard deviation of 10\%. This shows that the invariant is indeed conserved.

 \subsection{Expanding or contracting medium}
 \label{Sec_inv_expansion}

The derivation of Ref. \cite{Jaupart_GeneralizedTransportEquation2021a} is more general than the one of \cite{Chandrasekhar_FluctuationsDensityIsotropic1951} because it shows that \(M_{\rm inv}\) is still time invariant in a evolving background. The invariant should be conserved in a globally expanding or compressing medium in which \(\mathbb{E}(\rho)\) changes with time.

To show that this is indeed the case, our consideration on the polytropic equation of state with \(\gamma=5/3\) is interesting because it allows to deduce the behaviour of the invariant in an isotropically expanding or contracting box with velocity \(\bm{V}=H(t)\bm{r}\). This type of flow, called Hubble flow in the astrophysical context and widely used in cosmology to describe the expansion of the universe. Ref. \cite{Martel_ConvenientSetComoving1998} shows that the equations of conservation of mass, momentum and energy have the same form in a non-expanding medium with a polytropic equation of state with \(\gamma=5/3\) and in the comoving frame of an expanding medium if we choose the following transformation between the physical frame and the comoving frame:
\begin{align}
    \label{eq_transformation_comoving}
    \tilde{\bm{r}} &= a^{-1}\bm{r}, \nonumber \\
    {\rm d}\tilde{t} &= a^{-2} {\rm d}t, \\
    \tilde{\rho} &= a^3 \rho, \nonumber
\end{align}
where the tilde variables are those in the comoving frame and \(a\) is the expansion factor. It can be related to the Hubble constant \(H(t)\) by \(H(t)= \frac{1}{a}\frac{{\rm d}a}{{\rm d}t}\). In the comoving frame, \(\mathbb{E}(\tilde{\rho})\) is constant in time and the problem in this frame is thus equivalent to the one already solved in Sec. \ref{Sec_steady_state_back} {with the polytropic equation of state}.

Since the equations of conservation are the same for a polytropic gas with \(\gamma=5/3\) and in the comoving frame of an expanding/contracting flow, we also show from the polytropic simulation that the invariant is also verified in the comoving frame of an expanding/contracting medium. Thus we have:
\begin{equation}
    \tilde{M}_{\rm inv} = \mathbb{E}(\tilde{\rho})\text{Var}\left(\frac{\tilde{\rho}}{\mathbb{E}(\tilde{\rho})}\right)_t(\tilde{l}_{\rm c}^\rho)_t^3 = \text{const}.
\end{equation}
From the transformation given by Eq. \ref{eq_transformation_comoving} we have that \(M_{\rm inv} = \tilde{M}_{\rm inv}\). Thus the invariant is also verified in the physical frame.

\section{Analytical modelling of the correlation length and the invariant}
\label{Sec_Model}

\subsection{Gaussian model for the logdensity field: predicting \(l_{\rm c}^\rho\)}
\label{Sec_Gaussian_model}

\begin{figure}
    \centering
    \includegraphics[width=\columnwidth]{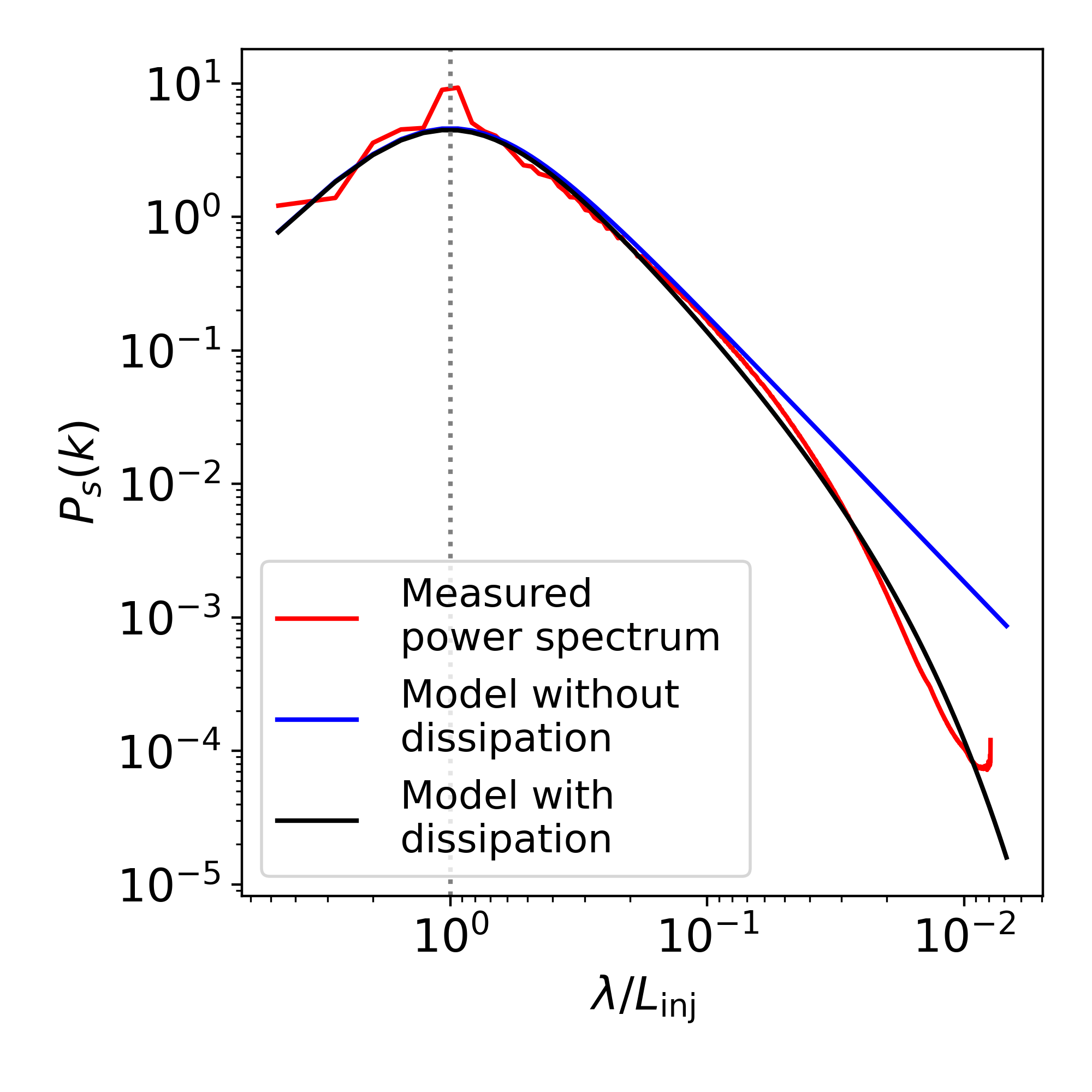}
    \caption{In red, the 1D power spectrum of the logdensity field measured during the stationary regime of the \(\mathcal{M}=3.5\) 1024\(^3\) simulation and the suggested model [see Eq. (\ref{eq_PS_logdensity_diss})] with the dissipation factor \(\beta=4\Delta x\) and without dissipation in blue (\(\beta=0\Delta x\)). They are both similar to each other, especially in the inertial range. The grey vertical dotted line shows the injection length.}
    \label{fig_PS_log_density}
 \end{figure}

We now present a model for predicting the evolution of the correlation length with the Mach number. We start with the first order assumption that the logdensity \(s=\ln(\rho/\mathbb{E}(\rho))\) is a 3D Gaussian random field. Whereas, strictly speaking,  there is no exact physical or mathematical justification for the assumption of Gaussianity for \(s\), a reasonable justification has been given in Ref. \cite{Dumond_ImpactShapePrestellar2024} for a gravo-turbulent medium.
The \(s\) field is completely characterized by the following power spectrum:
\begin{align}
    P_s^{\rm 1D}(k) \simeq A\frac{8\pi\sigma_s^2 L_{\rm inj}^3 \e^{-k\beta/(2\pi)}}{\left(1+\left(\frac{kL_{\rm inj}}{2\pi}\right)^2\right)^2}k^2,
    \label{eq_PS_logdensity_diss}
\end{align}
where \(k\) is the wave vector, $\sigma_s^2$ is the variance of the logdensity \(s\) and \(A\) is a coefficient close to unity that ensures that the integral of the power spectrum is equal to the variance of the logdensity field.
The coefficient \(\beta\) corresponds to the dissipation scale. The dissipation range is found to be universal in the grid simulations where the dissipation is induced by the grid \citep{Federrath_STARFORMATIONEFFICIENCY2013}. As a consequence, we chose an unique value of \(\beta\) normalized to the spatial resolution \(\Delta x\): \(\beta = 4 \Delta x\). The power spectrum given in Eq. (\ref{eq_PS_logdensity_diss}) is plotted in red in Fig. \ref{fig_PS_log_density}.

This form reproduces particularly well the one measured in our numerical simulations as shown in Fig. \ref{fig_PS_log_density}. The factor \(2\pi\) in Eq. (\ref{eq_PS_logdensity_diss}) can be understood as follows. In the code, the forcing is performed at the wavevector \(k_{\rm inj}=2\pi/L_{\rm inj}\) with \(L_{\rm inj} = L_{\rm box}/7\) in most of our simulations [see Eq. (\ref{eq_forcing_scale})].
It is quite natural that this particular scale is imprinted in some of the fields of the system. However, it is unclear why it affects in particular the logdensity field. This point should be investigated in more detail in dedicated studies.

With the two above assumptions, i.e. that the logdensity is a Gaussian random field and that its power spectrum has a form given by Eq. (\ref{eq_PS_logdensity_diss}), the statistics of the \(s\)-field is specified. While the ACF of the logdensity defined by Eq. (\ref{eq_def_ACF}) has no simple analytical form, we can still derive the correlation length of the logdensity field:
\begin{widetext}
\begin{equation}
    \label{eq_lc_s}
    l_{\rm c}^s=\frac{L_{\rm inj}}{2\pi}\left(\frac{\pi^2}{2 \text{Ci}(\tilde{\beta} ) (\sin (\tilde{\beta} )+\tilde{\beta}  \cos (\tilde{\beta} ))+(\pi -2 \text{Si}(\tilde{\beta} )) (\cos (\tilde{\beta} )-\tilde{\beta}  \sin (\tilde{\beta} ))}\right)^{1/3}
\end{equation}
\end{widetext}
where \({\rm Ci}\) and \({\rm Si}\) are respectively the cosine and sine integral function\footnote{The cosine and sine integral function are defined as follows: \(\text{Ci}(z)=-\int_z^{\infty } \frac{\cos (t)}{t}{\rm d}t\) and  \(\text{Si}(z)=\int_0^{z } \frac{\cos (t)}{t}{\rm d}t\).} and \(\tilde{\beta}=\beta/L_{\rm inj}\).

\begin{figure*}
    \centering
    \includegraphics[width=\textwidth]{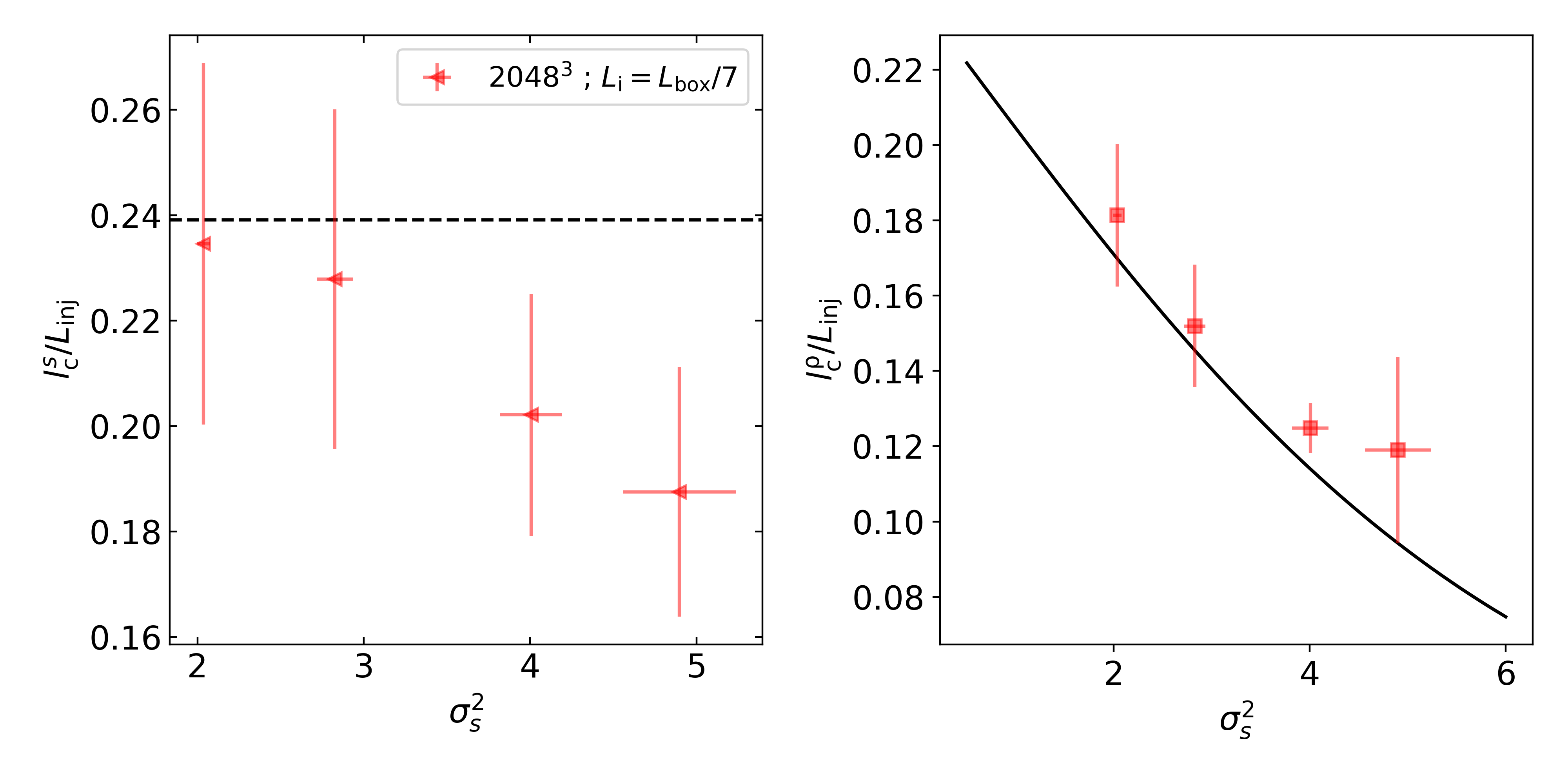}
    \caption{Evolution of the correlation length of the logdensity field (left) and the density field (right) with the variance \(\sigma_s^2\) of the logdensity field. The model for the logdensity correlation length is plotted as a dashed line [see Eq. (\ref{eq_lc_s})] in the left panel. The model for the density correlation length is plotted as a solid line [see. Eq. (\ref{eq_lc_rho})] in the right panel. The measurements of our $2048^3$ fiducial simulations are shown in red. The error bars correspond to the $\pm 1\sigma$ estimated from the time variations of the measured quantities.}
    \label{fig_lc_rho_sigma_Mach}
 \end{figure*}

From the Gaussian assumption on the \(s\)-field, we get the correlation of the normalized density field \(\tilde{\rho}=\rho/\mathbb{E}(\rho)\):
\begin{equation}
    \label{eq_corr_rho}
    C_{\tilde{\rho}}(\bm{q})=\exp(C_{s}(\bm{q}))-1.
\end{equation}
This expression enables us to compute the correlation length of the density field:
\begin{align}
    l_{\rm c}^{\rho} &= \left(\frac{\pi}{2}\frac{1}{C_{\tilde{\rho}}(0)}\int_{0}^{\infty}C_{\tilde{\rho}}(q)q^2{\rm d}q\right)^{1/3}.
    \label{eq_lc_rho}
\end{align}
This integral has no analytical expression.
The variation of the density correlation length with \(\sigma_s\) predicted by the model is plotted as a solid line in the right panel of Fig. \ref{fig_lc_rho_sigma_Mach}. This model only holds when the turbulence is forced and fully developed. It does not give a prediction for the evolution of the correlation length of the density field when the turbulence is decaying, as the power spectrum of the log density field would not have the functional form given by Eq. (\ref{eq_PS_logdensity_diss}) anymore. Therefore, the evolution of the correlation length of the log density field is not known.

We now compare the prediction of our model with the measurements of the correlation lengths made in the simulations after reaching the statistically stationary regime and before turbulence starts to decay. In Fig. \ref{fig_lc_rho_sigma_Mach} we have plotted the average correlation length of the density and logdensity fields, respectively, measured in the four 2048\(^3\) runs with turbulence forced at scale \(L_{\rm box}/7\) and Mach numbers ${\cal M}$=3.5, 5, 7 and 10. We did not show the correlation length measured at \(\mathcal{M}=2.5\) because it is not converged relative the size of the box  (see Appendix \ref{App_num_convergence}). The error bars correspond to the uncertainties at \(1\sigma\) estimated from the time variations of the measured quantities. As can be seen in the figure, our model agrees well (less than one \(\sigma\)) with the measured correlation lengths of the density and the logdensity fields {\it without any fitting parameters}. Some discrepancies (around 1.5 - 2 \(\sigma\)) between the measurements and the model are observed at high Mach, especially for the logdensity correlation length. We found that this is due to the flattening of the logdensity power spectrum with Mach number, which is not taken into account in the model, but also to the fact that non-Gaussian features of the logdensity field become non neglegible at high Mach.
The convergence studies of the density and logdensity correlation lengths are presented in Appendix \ref{App_num_convergence}.

\subsection{Modelling the evolution of \(M_{\rm inv}\) with the Mach number}
\label{Sec_M_inv_model}
\begin{figure}
    \centering
    \includegraphics[width=\columnwidth]{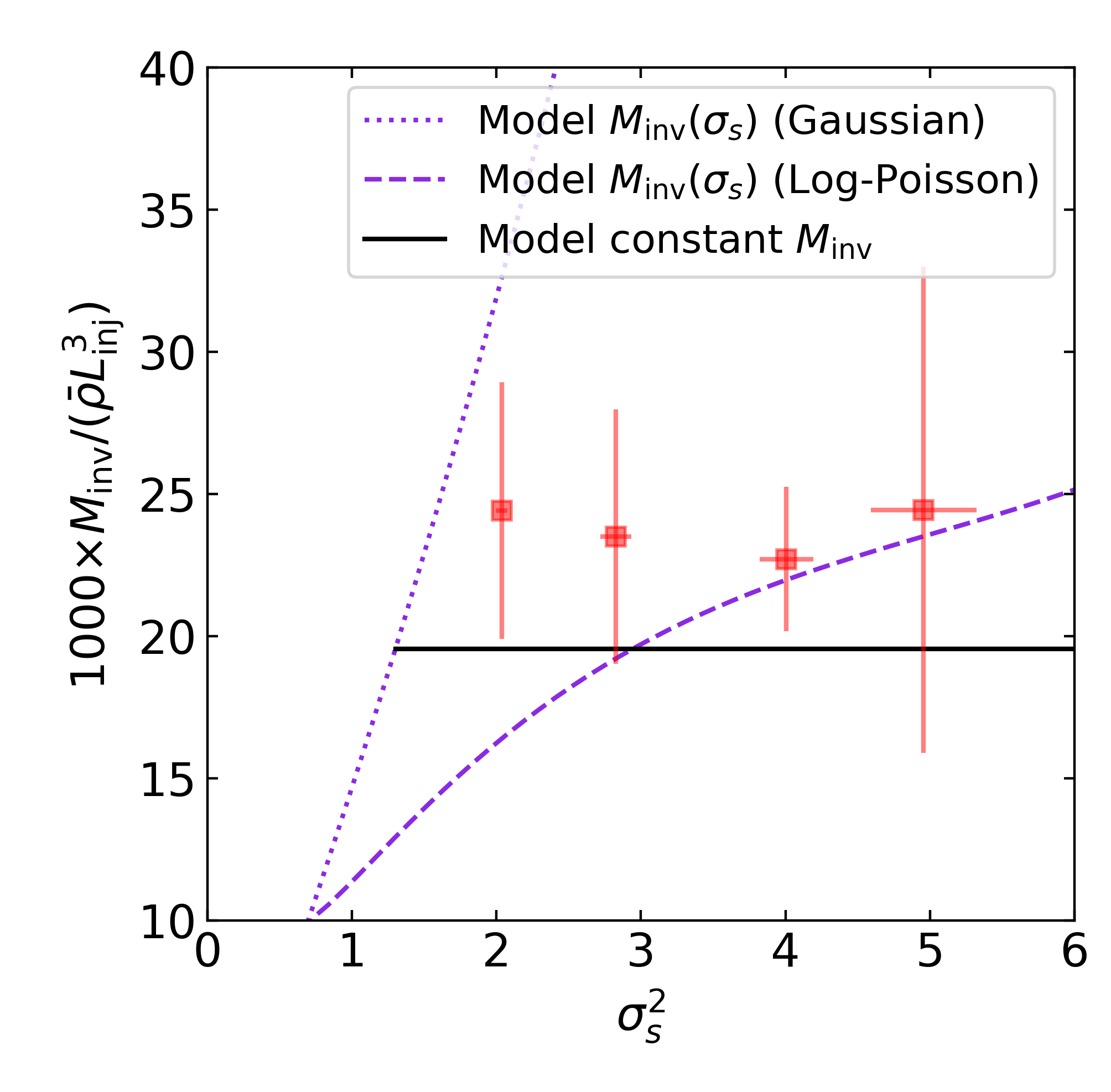}
    \caption{Evolution of the invariant \(M_{\rm inv}\) with the variance \(\sigma_s^2\) of the log density field. The model for the mass invariant based on the Gaussian assumption only is plotted as a purple dotted line. The model with a non-Gaussian correction for the density variance is plotted as a purple dashed line. Both rely on the model of the correlation length described in Sec. \ref{Sec_Gaussian_model}. The model describing a constant mass invariant is plotted in black solid line (see Sec. \ref{Sec_Intermittency}). The measurements from a 2048\(^3\) simulation with turbulence injected at \(L_{\rm box}/7\) are shown in red. The error bars correspond to the $\pm 1\sigma$ estimated from the time variations of the measured quantities.}
    \label{fig_M_inv_sigma_Mach}
 \end{figure}

From the model of the correlation length presented in the previous section, we can model the dependence of \(M_{\rm inv}\) on the Mach number during the statistically stationary regime. Within the Gaussian approach, its expression can be derived as:
\begin{multline}
    \label{eq_mass_invariant}
    M_{\rm inv}= \mathbb{E}(\rho)L_{\rm inj}^3 \frac{\pi}{2 C_{\tilde{\rho}}(0)} \times \\  \int_0^\infty C_{\tilde{\rho}}(\tilde{q}L_{\rm inj}) \tilde{q}^2{\rm d}\tilde{q}  \left(\e^{\sigma_s^2}-1\right) .
\end{multline}
This model is plotted in purple dotted line in Fig. \ref{fig_M_inv_sigma_Mach}. We see that this model does not reproduce at all the measurements of \(M_{\rm inv}\) made in the simulations after reaching the statistically stationary regime. This reflects the fact that the variance \(\sigma_{\tilde{\rho}}^2=b^2\mathcal{M}^2\) of the density field predicted from the variance of $s$ under a Gaussian field assumption is significantly overestimated. Consequently, Eq. (\ref{eq_mass_invariant}) overestimates \(M_{\rm inv}\), especially at high Mach.

The model proposed above has the major drawback of relying on the Gaussian assumption for the statistics of the logdensity random field. It is well known that although the PDF of the logdensity field can be reasonably well modelled by a Gaussian function, the non-Gaussianities of turbulence  reduces its high-density tail, thus its variance \citep{Federrath_ComparingStatisticsInterstellar2010,Hopkins_ModelNonlognormalDensity2013a}.
To better predict the evolution of the mass invariant with Mach number, we use the log-Poisson model to describe the logdensity PDF \citep{Dubrulle_IntermittencyFullyDeveloped1994, Castaing_TemperatureTurbulentFlows1996, Hopkins_ModelNonlognormalDensity2013a,Squire_DistributionDensitySupersonic2017} (see also Ref.\cite{Mocz_MarkovModelNonlognormal2019} for another model). The logdensity PDF has thus the following form:
\begin{equation}
        \label{Log_poisson}
        \mathcal{P}(s)=I_1(2 \sqrt{\lambda u}) \e^{-(\lambda+u)} \sqrt{\frac{\lambda}{T^2 u}},
\end{equation}
defining the quantities
\begin{equation}
    u = \frac{\lambda}{1+T}-\frac{s}{T}, \  \lambda=\frac{\sigma_s^2}{2 T^2}.
\end{equation}
This introduces a Mach-dependent parameter \(T\) to reproduce the negative skewness of the density PDF observed in compressible turbulence simulations: the larger \(T\) the larger the negative skewness. In this model, the variance of the density field becomes:
\begin{equation}
    \sigma_{\tilde{\rho}}^2 = \exp\left(\frac{\sigma_s^2}{1+3T+2T^2}\right)-1.
\end{equation}
For \(T=0\), we recover the Gaussian prediction for \(\sigma_{\tilde{\rho}}\). From a phenomenological model of the structure of supersonic shocks, a parametrization of the dependence of \(T\) on the Mach number as \(T=\kappa(1-(b\mathcal{M})^{-2})\) has been proposed \cite{Squire_DistributionDensitySupersonic2017}. We find that a value \(\kappa=0.16\) best fits our data, slightly different from \(\kappa=0.2\) found by the aforementioned authors. The Mach number can be related to the logdensity variance by the following equation:
\begin{equation}
    \label{eq_sigma_s_Mach}
    \sigma_s^2=\ln\left(1+b^2\mathcal{M}^2\right),
\end{equation}
with a forcing parameter \(b\) whose value depends on the nature of the turbulent forcing \citep{Federrath_ComparingStatisticsInterstellar2010,Molina_DensityVarianceMachNumber2012a,Konstandin_MachNumberStudy2016}. These studies suggest that for mixed driving (solenoidal fraction \(\zeta=0.5\)), \(b\) should be close to 0.4. This value differs, however, from other studies \citep{Brucy_InefficientStarFormation2024}. Here we find that \(b=0.8\) reproduces best the relation we measure between \(\sigma_s^2\) and ${\cal M}$.

The expression of the invariant then becomes:
\begin{multline}
    \label{eq_mass_invariant_non_gauss}
    M_{\rm inv}= \mathbb{E}(\rho)L_{\rm inj}^3 \frac{\pi}{2 C_{\tilde{\rho}}(0)} \times \\ \int_0^\infty C_{\tilde{\rho}}(\tilde{q}L_{\rm inj}) \tilde{q}^2{\rm d}\tilde{q} \left(\exp\left(\frac{\sigma_s^2}{1+3T+2T^2}\right)-1\right) .
\end{multline}
This non-Gaussian correction concerns only the relation between the density and the logdensity variance, not the one between the density and the logdensity correlation lengths, which seems to be well described by the Gaussian model proposed in Sec. \ref{Sec_Gaussian_model}. Taking into account the non-Gaussianities in the relationship between the ACFs of the logdensity and density field from a theoretical point of view is much more difficult because it would require the knowledge of all the orders of the two points correlation of the logdensity field.

In Fig. \ref{fig_M_inv_sigma_Mach}, we plot the mass invariant against \(\sigma_s^2\). Our model, which includes the above correction for the non-Gaussianity of the logdensity PDF, reproduces the measurements reasonably well at high Mach, while it significantly underestimates the value at low Mach.

\subsection{Discussion of the hypotheses}
\label{Sec_Discussion}

As described above, our basic Gaussian model relies on a first order assumption about the statistics of the logdensity field, namely that it is a Gaussian random field. This assumption is inaccurate because the supersonic shocks generate high order correlations, which induce phase coupling within the field (see Ref. \cite{Dumond_ImpactShapePrestellar2024} for a discussion) stronger at high Mach. Therefore, the statistics of \(s\)-field cannot be described only by its power spectrum and PDF. However, the fact that our Gaussian model successfully reproduces the evolution of the density correlation length, \(l_{\rm c}^{\rho}\), with the variance of the logdensity field, \(\sigma_{s}^2\), suggests that the underlying assumptions describe the statistics of the logdensity field sufficiently well.

\begin{figure*}
    \centering
    \includegraphics[width=\textwidth]{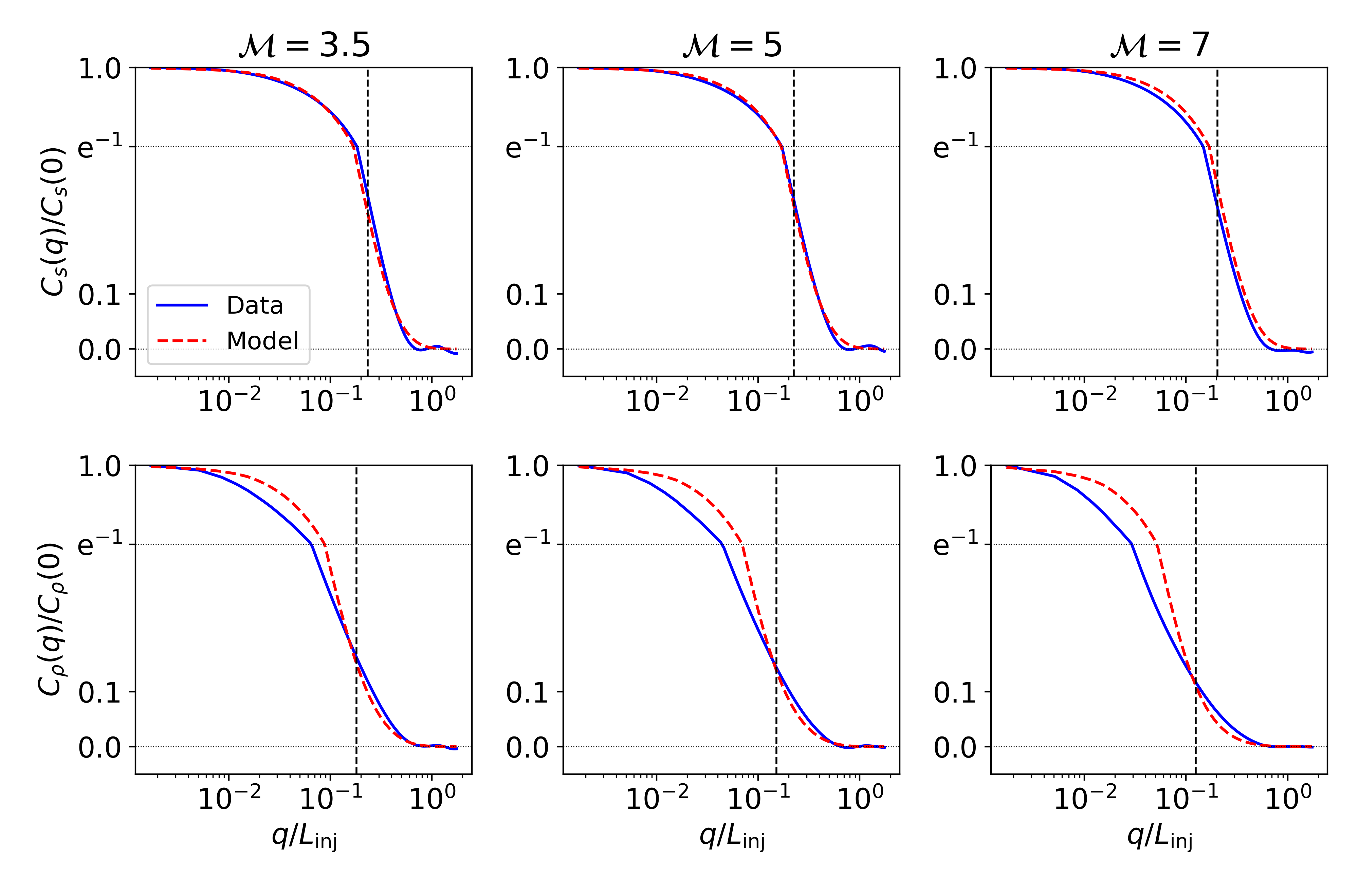}
    \caption{Autocovariance function of logdensity \(s\) (top) and density \(\rho\),(bottom) normalised to their respective variance measured in the \({\cal M}\)=2.5, 3.5 and 7 runs during the statistically stationary phase. Our model for the logdensity and density autocorrelation functions is shown by the red dashed line. The measured value of the correlation length of the logdensity and density fields is shown by the vertical black dashed line. The \(y\)-scale is log above \(\e^{-1}\) and linear below.
    }
    \label{fig_autocorr_func}
 \end{figure*}

In Fig. \ref{fig_autocorr_func} we compare the predicted ACFs of the density and logdensity fields, respectively, with the ones measured in the three 2048\(^3\) runs at \({\cal M}=\)2.5, 3.5 and 7. As seen, there is good agreement, without any fitting parameters. As the Mach number increases, Eq. (\ref{eq_corr_rho}) becomes increasingly inaccurate to describe the two-point statistics of the density field.
It must be kept in mind that the non-Gaussian correction to the model (see Sec. \ref{Sec_M_inv_model})  modifies only the variance of the density field but not the relation between the density and logdensity ACFs
[see Eq. (\ref{eq_corr_rho})]. Thus, as shown in Fig. \ref{fig_autocorr_func}, the functional form of the density field ACF predicted by Eq. (\ref{eq_corr_rho}) deviates from the one obtained in the simulations at high \({\cal M} \) because of non-Gaussian processes. However, this deviation has small consequences on the accuracy of the model to predict the correlation length (see Fig. \ref{fig_lc_rho_sigma_Mach}) because the value of the integral of the ACF normalized to the density variance is not very sensitive to these small fluctuations of its functional form, at least for \({\cal M}\lesssim \)10. More precisely, we can note that the deviation of the model compared to the measured one is large at short distance, but much smaller at large one. This expected since the non Gaussianities will play an important role at short distance when the structures are highly correlated. They will be less important at large distance, where the fields are almost independent. Because the correlation length involves the integral of \(q^2 C_\rho(q)\), the discrepencies observed between the model and the data plays a minor role in the result of the integral.
This explains the success of the model in describing the evolution of the correlation lengths as the function of \(\sigma_s\) and \({\cal M}\).

\section{Derivation of the law \(T(\sigma_s)\)}
\label{Sec_Intermittency}

As an example of what we can learn from this invariant, we show that we can compute the variation of the parameter \(T\) with the Mach number, and thus predict some of the non-Gaussian properties of the density field.

\subsection{Dependence of the Mass Invariant \(M_{\rm inv}\) with the Mach number}
\label{Sec_independent_Mach_inv}

In this subsection we first give arguments showing that the quantity \(M_{\rm inv}\) should be independent of the Mach number. Based on this, in the next subsection, we will derive the evolution of the logdensity PDF with \(\sigma_s\) and \(\mathcal{M}\) and compare it with the existing parameterisations of this relation in the literature.

Let's consider a medium that is initially at rest, in which turbulence is driven. Once the correlation length is small enough compared to the box size to be properly resolved, the mass \(M_{\rm inv}\) is a constant quantity. In particular, when we stop driving the turbulence and let it decay, this quantity remains constant. It only breaks down when the correlation length is too large compared to the size of the system, i.e. when volumic averages are bad estimates of statistical averages. Instead of letting the turbulence decay, we can go the other way and increase the driving force until a new target Mach number is reached. During this process, none of the hypotheses on which the invariant is based is violated. The quantity \(M_{\rm inv}\) should therefore remain constant during this process. Since this can be done for any Mach number, we conclude that the invariant should be constant for any Mach number.

However, by considering the limit \(\mathcal{M}\to 0\) we can see that this reasoning is limited. In fact, at very low Mach, the variance of the density field tends to 0, while the correlation length cannot go to infinity because it is limited by the injection length. Indeed, regions of the fluid separated by distances larger than injection length are completely uncorrelated. This can be seen from the shape of the 3D power spectrum of the density which is flat for \(k<k_{\rm inj}\) with \(k_{\rm inj}=2\pi/L_{\rm inj}\).
Thus, at least in forced subsonic turbulence, the invariant should depend on the Mach number. This dependence on the Mach number in the subsonic regime is the consequence of the fact that the cross-correlation between the density and the momentum is not negligible due to the large scale correlations imposed on the flow by the driving. Once we reach the supersonic regime, the cross-correlation becomes negligible.

\begin{figure}
    \centering
    \includegraphics[width=\columnwidth]{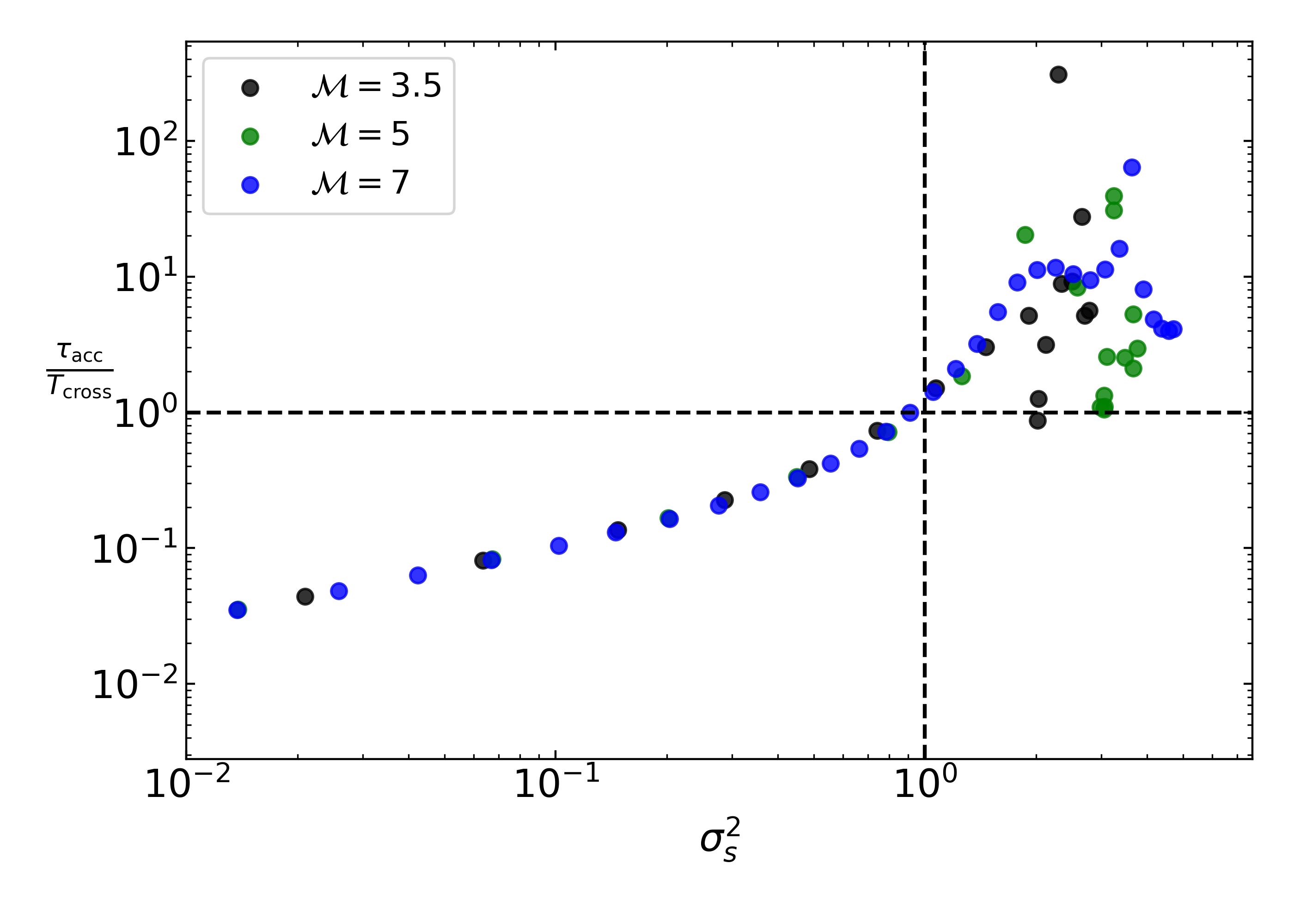}
    \caption{The black dots shows the measurement of \(\tau_{\rm acc}/T_{\rm cross}\) in the 1024\(^3\) run during the transitory regime for different target Mach numbers: \(\mathcal{M}=3.5\) (black), \(\mathcal{M}=5\) (green) and \(\mathcal{M}=7\) (blue). It is evaluated at the first zeros of the ACF \(\mathcal{C}_\rho\).
    The ratio is small in the subsonic regime (\(\sigma_s^2\lesssim 1\)), suggesting that the invariant in this regime is strongly dependent on the Mach number. In the supersonic regime (\(\sigma_s^2\gtrsim 1\)) it increases significantly, confirming that the invariant will be independent of the Mach number in this regime.}
    \label{fig_condition_no_acc}
\end{figure}

This behaviour can be seen by considering the accretion timescale \(\tau_{\rm acc}\) of the invariant, corresponding to the time needed to modify it substantially. From the computation of the invariant derived by \cite{Chandrasekhar_FluctuationsDensityIsotropic1951} and \cite{Jaupart_GeneralizedTransportEquation2021a}, it is defined as
\begin{equation}
    \tau_{\rm acc} = \Bigg\lvert \frac{\int_\Omega C_\rho(q){\rm d}^3q}{\int_{\partial\Omega} \mathbb{E}(\rho \rho' \bm{v}'){\rm d}^2\bm{S}} \Bigg\rvert
\end{equation}
where \({\rm d}^2\bm{S}\) is an infinitesimal surface element and where \(\Omega\) is the measurement domain, typically the sub-box of size \(L_{\rm box}/2\).. To ensure that the invariant is constant, the accretion time \(\tau_{\rm acc}\) should be greater than the turbulent crossing time \(T_{\rm cross}=L_{\rm inj}/\sqrt{\mathbb{E}(\bm{v}^2)}\).
In Fig. \ref{fig_condition_no_acc}, we plot the measurement of \(\tau_{\rm acc}\) done in the 1024\(^3\) run divided by the crossing time during the transitory regime. We do not perform this calculation on a 2048\(^3\) simulation because the high computational cost of this calculation, but to study a 1024\(^3\) simulation is sufficient given the convergence study (see Appendix \ref{App_num_convergence}). \(\Omega\) is taken to be a ball whose radius is the lag corresponding to the first zeros of the ACF.
We see that these measurements cross the horizontial line \(\tau_{\rm acc}/T_{\rm cross}=1\) at \(\sigma_s^2\simeq 1\). \(M_{\rm inv}\) should thus be constant soon after the time such that \(\sigma_s^2=1\), i.e. in the supersonic regime, as observed in the simulation.

To compute the value of the invariant, we focus on the transonic regime where the statistical properties of the density field are reasonably well known. For a Mach number close to 1, the density PDF is well approximated by a lognormal (e.g. \cite{Hopkins_ModelNonlognormalDensity2013a}), i.e \(T\ll 1\). Basicaly, we want to choose the largest supersonic Mach number (or largest \(\sigma_s^2\)) at which \(T\) is still small, typically below a few percent, such that the \(s\)-PDF is still well approximated by a Gaussian and not to be affected by the variation of the invariant in the subsonic regime.
In Eq. (\ref{eq_mass_invariant}), we choose \(\sigma_s^2 = 1.3\) because it corresponds to a Mach number of 2. This choice is motivated by the fact that at such Mach number, the measurements of \(T\) available in the literature report a value between 1\% and 5\%. Moreover, \(\sigma_s^2>1\) suggests from Fig. \ref{fig_condition_no_acc} that \(M_{\rm inv}\) should already be constant.
With this value, we find that the mass of the invariant is equal to \(\alpha M_{\rm inj}\) with \(\alpha=2.0\times 10^{-2}\) and \(M_{\rm inj}=\mathbb{E}(\rho)L_{\rm inj}^3\). This corresponds to the the solid black horizontal line in Fig. \ref{fig_M_inv_sigma_Mach}. The fact that the invariant is constant for different Mach numbers is consistent with the numerical data plotted in Fig. \ref{fig_M_inv_sigma_Mach}.

Our analytical model slightly underestimates its value because the model of the correlation length also slightly underestimates the measured one and \(M_{\rm inv}\) depends steeply on \(l_{\rm c}^\rho\). This is a consequence of the non-Gaussianities not taken into account by the model.

\subsection{Computation of the parameter \(T\)}

In the framework described above, we can determine the evolution of the parameter \(T\) introduced by \cite{Hopkins_ModelNonlognormalDensity2013a} for compressible turbulence versus \(\sigma_s\) by solving the following equation derived from Eq. (\ref{eq_mass_invariant_non_gauss}) using \(M_{\rm inv}=\alpha\mathbb{E}(\rho)L_{\rm inj}^3\)
\begin{multline}
    \label{eq_T_sigma_s}
    \alpha= \frac{\pi}{2 C_{\tilde{\rho}}(0)} \int_0^\infty (\exp(C_{s}(\tilde{q} L_{\rm inj}))-1) \tilde{q}^2{\rm d}\tilde{q} \times \\
    \left(\exp\left(\frac{\sigma_s^2}{1+3T+2T^2}\right)-1\right) ,
\end{multline}
where \(\alpha\) is the constant determined in the previous section. The determination of the parameter \(T\) by this method assumes that the previously derived model for the density correlation length is accurate even at high Mach numbers. This is confirmed by our numerical simulations, at least for Mach numbers below 8 (see Sec. \ref{Sec_Model}).

\begin{figure*}
    \centering
    \includegraphics[width=\textwidth]{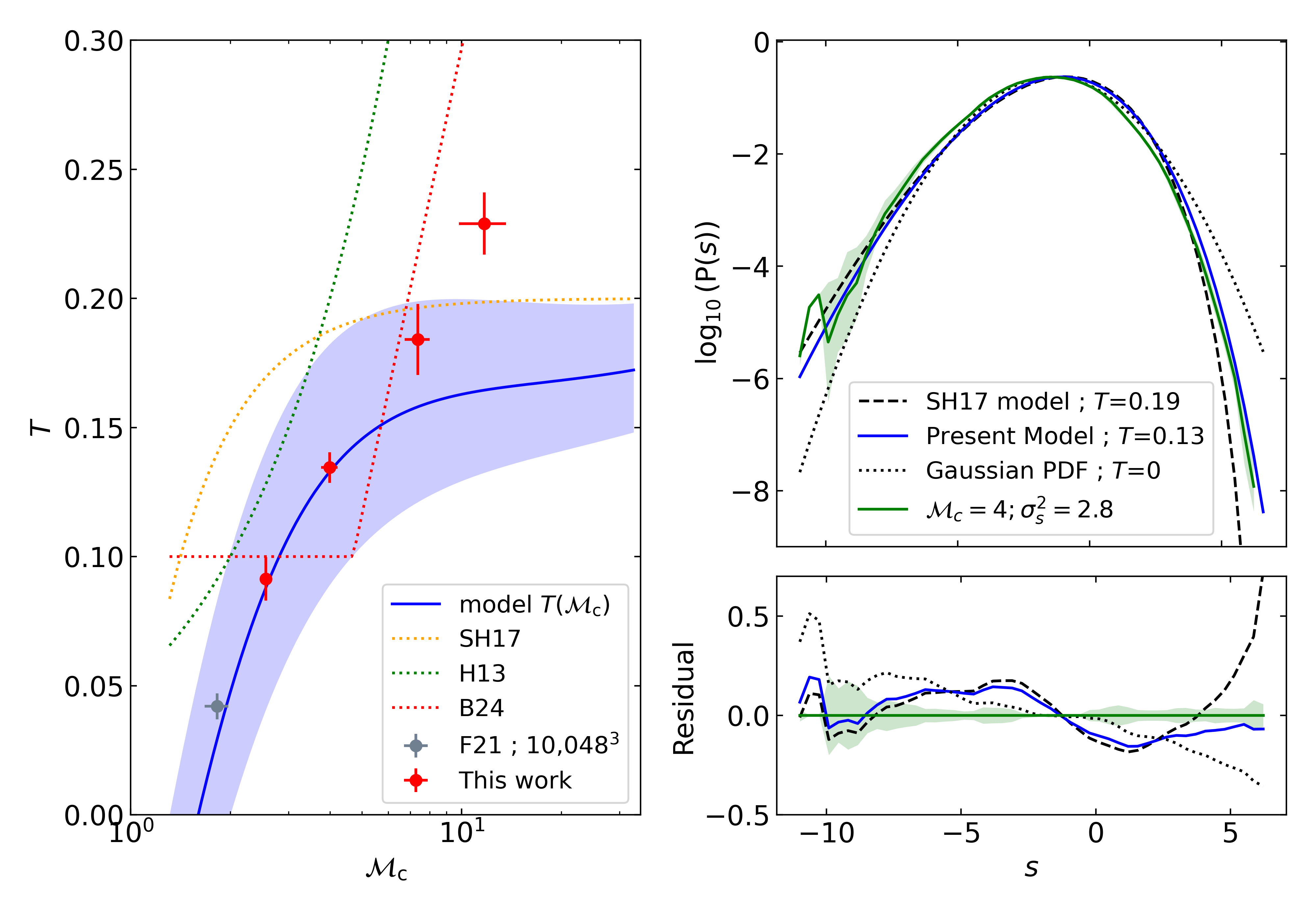}
    \caption{Left: Evolution of the parameter \(T\) with the compressive Mach number \(\mathcal{M}_{\rm c}\). Different parametrizations of \(T\) found in the literature are shown in dotted green \citep{Hopkins_ModelNonlognormalDensity2013a}, orange \citep{Squire_DistributionDensitySupersonic2017} and red \cite{Brucy_InefficientStarFormation2024}. The red points corresponds to the \(T\) parameter extracted by fitting the measured PDF with the fonctional form suggested by \cite{Hopkins_ModelNonlognormalDensity2013a} (his eq. 5) in our four \(2048^3\) simulations with turbulence forced at \(L_{\rm box}/7\). The gray point is the measurement from the high resolution turbulence simulation from \cite{Federrath_SonicScaleInterstellar2021}. The error bars correspond to the $\pm 1\sigma$ estimated from the time variations of the measured quantities. The model is shown in blue. The shaded area corresponds to different prediction of the model considering various values of the invariant. Top right: In green, measured \(s\)-PDF in the 2048\(^3\) simulation at \(\mathcal{M}=5\). The shaded area corresponds to the 1\(\sigma\) uncertainties estimated from the time variations of the PDF. The blue solid line is the PDF predicted by the model presented in Sec.\ref{Sec_Intermittency}. The dotted black line corresponds to the lognormal mode of the PDF (\(T=0\)) and the dashed line corresponds to the PDF predicted by \cite{Squire_DistributionDensitySupersonic2017}. Bottom right: Relative error between each model plotted on the top panel and the measured PDF. }
    \label{fig_T_sigma_S_Mach}
\end{figure*}

To predict the variation of \(T\) with the Mach number instead of the logdensity variance \(\sigma_s^2\), we use Eq. \ref{eq_sigma_s_Mach}.
In the left panel of Fig. \ref{fig_T_sigma_S_Mach} we plot the evolution of \(T\) with the Mach number. They are obtained by fitting the \(s\)-PDF measured in our simulation with the log-Poisson PDF (see Eq. \ref{Log_poisson}) with \(T\) being the only fit parameter. As seen in most of the numerical simulations, we see that this parameter increases with the Mach number. We compare this with other laws for \(T\) derived from numerical simulations and/or phenomenological models \citep{Hopkins_ModelNonlognormalDensity2013a, Squire_DistributionDensitySupersonic2017,Brucy_InefficientStarFormation2024}. All these models are parameterized with the compressive Mach number \(\mathcal{M}_{\rm c} = b\mathcal{M}\).

The model derived with the value of the invariant computed at Mach 2 is ploted in blue solid line. Because this value is subject to uncertainties, the shaded area corresponds to the prediction of the model with an invariant calculated at a Mach between 1.6 and 2.5, corresponding to \(\sigma_s^2\) varying between 1 and 1.6. The best fit to our simulations, inc. that of F21 and excluding \(\mathcal{M}=10\) (see below), is obtained for \(\mathcal{M}=2\). The model reproduces well the measurements of \(T\) in the high resolution simulation performed by Ref. \cite{Federrath_SonicScaleInterstellar2021}, and the measurements in our 2048\(^3\) simulations for Mach numbers below 7 while the existing parametrisations in the literature overestimate it \citep{Hopkins_ModelNonlognormalDensity2013a,Squire_DistributionDensitySupersonic2017,Brucy_InefficientStarFormation2024}. This may suggests that the measurements of \(T\) in some of the low resolution numerical simulations are not well converged even at small Mach number because of a lack of statistics, the turbulence being usally injected at \(L_{\rm box}/2\). The parametrization given by \cite{Hopkins_ModelNonlognormalDensity2013a, Brucy_InefficientStarFormation2024} are also approximate, given the large scatter in the measurements of \(T\) used in these studies to determine these parametrizations. For Mach number larger than 7, the underestimation of \(T\) predicted by the model compared to our measurement in the Mach 10 simulation and to the existing parametrizations is likely to be a consequence of the non-Gaussianities that becomes dominant at such high Mach numbers. These non-Gaussian features result in an underestimation of the correlation length by the model described in Sec. \ref{Sec_Gaussian_model}. The range of validity of our model is therefore limited to Mach numbers between 2 and 7. Below Mach 2, \(M_{\rm inv}\) is not constant while above Mach 7, non Gaussian features are not negligible.

On the right panel, we show the \(s\)-PDF measured in our 2048\(^3\) simulation at \(\mathcal{M}=5\) and plot the PDF predicted by our model. The model reproduces very well the data. For comparison, we also plot the PDF that would be given by the Gaussian model, i.e. with \(T=0\). It significantly overestimates the amount of dense gas, highlighting the importance of considering a model able to describe the negative skewness of the \(s\)-PDF. To predict accurately the amount of dense gas in a supersonic turbulent medium is indeed of prime importance in an astrophysical context, namely to model the star formation process \citep{Brucy_InefficientStarFormation2024, Hennebelle_InefficientStarFormation2024}

\section{Conclusion}

In this article, we have  studied numerically the accuracy of the mass invariant introduced by Ref. \cite{Chandrasekhar_FluctuationsDensityIsotropic1951} for compressible isotropic turbulence and extended by Ref. \cite{Jaupart_GeneralizedTransportEquation2021a}. We show that the predicted time invariance of the quantity \(M_{\rm inv}\) is well verified during the decaying phase of isotropic turbulence. A correlation length small enough compared to the box size is crucial for a correct computation of the invariant. In order to fulfil this condition, turbulence must be injected at a scale significantly smaller than the size of the system under study. Although the calculations of Ref. \cite{Jaupart_GeneralizedTransportEquation2021a} for the invariant are valid for both isotropic and non-isotropic turbulence, we have studied here only the isotropic case which is representative of a vast variety of turbulent flows both in physics and astrophysics.
Anisotropic turbulence does not present any particular difficulties, but is outside the scope of this paper and will be considered in a future study.

Even though we use the Gaussian random field assumption which is a  first order assumption to describe the statistical properties of the supersonic flow, the derived model reproduces well the evolution of the correlation length $l_{\rm c}^\rho$ with $\cal M$. This suggests that the Gaussian random field assumption describes the statistics of the logdensity field sufficiently well.
However, in order to correctly describe the evolution of \(M_{\rm inv}\), we have derived analytically a non-Gaussian correction to better capture the evolution of the density variance $\sigma_{\tilde{\rho}}^2$ with $\cal M$. This improved model better agrees with the results of the simulations. Finally, we argue why \(M_{\rm inv}\) should not depend on the Mach number from the trans to supersonic regime (\(\mathcal{M}\sim 7\)) and show that this invariant can be a powerful tool to predict the evolution of the logdensity PDF with the Mach number with no free parameters. This is quite a fundamental result for our understanding of compressible turbulence as it provides an exact way to infer the evolution of the main statistical properties of a compressible turbulent flow.

In the astrophysical context, gravity plays a major role in the formation of structures in gravo-turbulent molecular clouds \citep{Vazquez-Semadeni_GlobalHierarchicalCollapse2019a, Hennebelle_PhysicalOriginStellar2024}. It modifies the statistical properties of the flow, as seen in the form of the density PDF, which develops a power-law tail at high density \citep{Jaupart_EvolutionDensityPDF2020, Schneider_UnderstandingStarFormation2022a}.
In that context, the invariant may represent a fundamental quantity for the formation of structures in molecular clouds, as it would determine the characteristic mass of the most correlated structures induced by turbulence in a gravitational system. On the theoretical point of view, this has been studied by \cite{Jaupart_GeneralizedTransportEquation2021a}. It will be thus of prime importance to verify the validity of this invariant and its physical interpretation in such a context, which will the purpose of a forthcoming study.

\begin{acknowledgments}
    The authors thank the referees for helpful comments that improved the clarity of the manuscript. The authors thank Laurent Chevillard for reading of the manuscript and his helpful comments. They also thank Quentin Vigneron, Elliot Lynch, Thomas Gillet, Noé Brucy, Michaël Bourgoin, Benoît Commerçon, Léa Cherry for helpful discussions and comments. Simulations were produced using GENCI allocations (grants A0150411111 and A0170411111). We gratefully acknowledge support from the CBPsmn (PSMN, Pôle Scientifique de Modélisation Numérique) of the ENS de Lyon for the computing resources. The platform operates the SIDUS solution \citep{Quemener_SIDUS} developed by Emmanuel Quemener. PD and JF were supported by the French national research agency grant ANR-23-CE31-0005 (BRIDGES).
\end{acknowledgments}

\appendix

\section{Numerical convergence}
\label{App_num_convergence}

In this appendix we present our convergence test for the different quantities of interest, namely the density and log density variances and the correlation lengths. First, we have increased the resolution of the simulation from 512\(^3\) to \(2048^3\) for the simulation where turbulence is forced at \(L_{\rm box}/7\).
The variances and correlation length are well converged for the \(2048^3\) simulation. Doubling the number of cells from 1024\(^3\) to \(2048^3\) changes the correlation length by less than 5\%. This corresponds to the case where the correlation is resolved by more than 10 cells. At high Mach (Mach 7 and 10) the logdensity variance shows a slightly larger variation, until 15\% when the resolution is doubled.
To properly confirm the convergence of the logdensity variance at these Mach numbers, a \(4096^3\) should be performed, but this requires too much computational resources.

To test the effect of the finite volume of the studied system, we also increased the size of the sub-box relative to the injection length. Therefore, we have performed 1024\(^3\) and 2048\(^3\) simulations where the turbulence is forced at \(L_{\rm box}/14\). The physical spatial resolution of this simulation is the same as in the \(512^3\) and \(1024^3\) run with turbulence forced at \(L_{\rm box}/7\) respectively.
When we force the turbulence at \(L_{\rm box}/14\) with a resolution of 1024\(^3\) at \(\mathcal{M}=3.5\), the logdensity variance and the density correlation length are similar to those measured in the 512\(^3\) simulation with turbulence forced at Mach \(L_{\rm box}/7\). This confirms that the size of the system has little effect on these quantities, provided it is large enough. However, at \(\mathcal{M}=2.5\) the density correlation length is not converged at all with respect to the injection length. This suggests that at such a small Mach number one needs a very large box to properly resolve the correlation length of the fields of interest. To do this, one should perform a \(4096^3\) simulation with an injection of turbulence at \(L_{\rm box}/28\) to confirm the convergence of the simulations run at \(L_{\rm box}/14\), but this also requires too much computational resource. This problem highlights the difficulty of this study, which requires both a very good spatial resolution to properly resolve small scale turbulent structures, and a turbulent box large enough compared to the injection scale of turbulence to have sufficient statistics.

\begin{figure*}
    \centering
    \includegraphics[width=\textwidth]{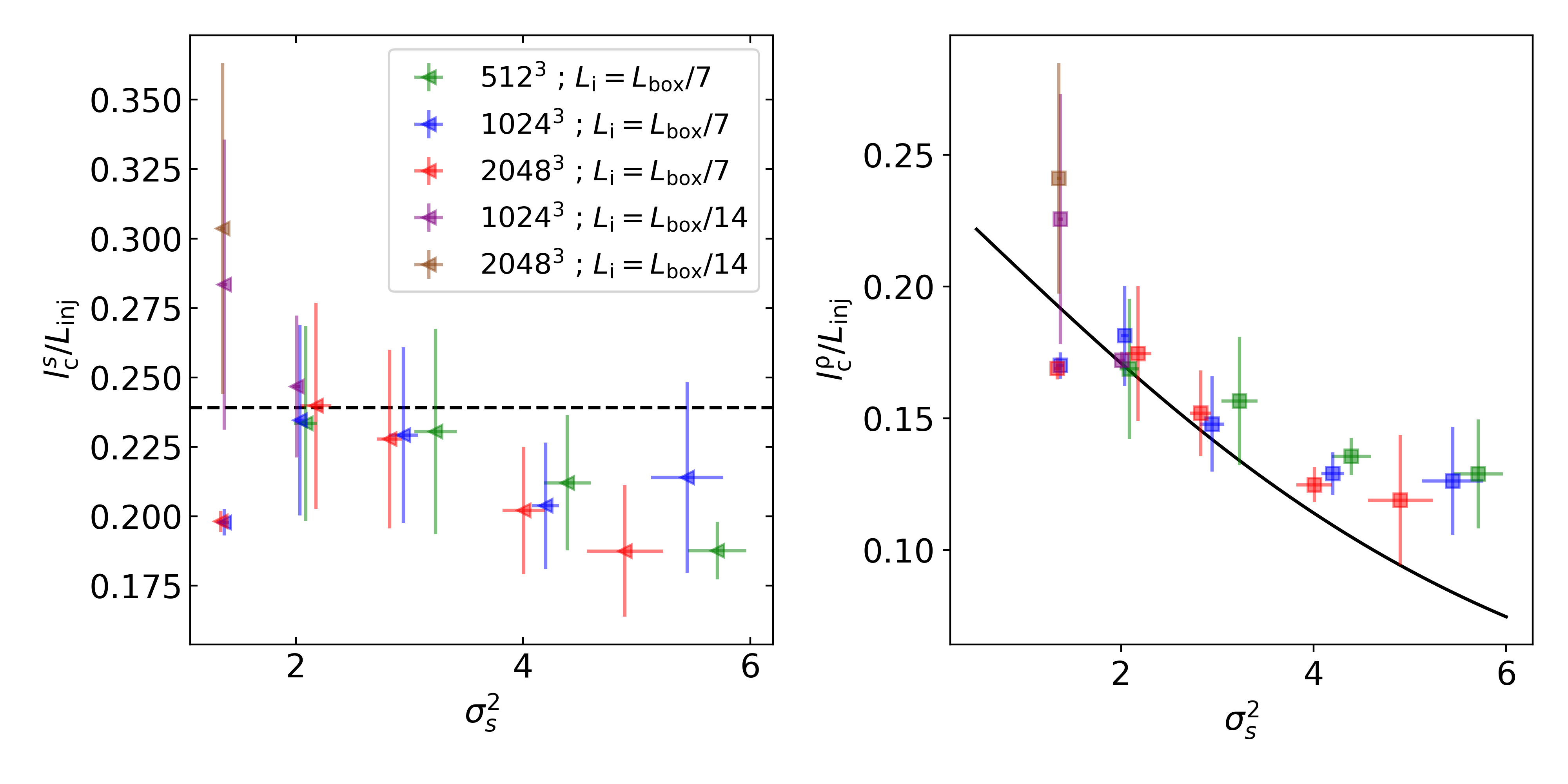}
    \caption{Evolution of the correlation length of the logdensity field (left) and the density field (right) with the variance \(\sigma_s^2\) of the logdensity field. The model for the logdensity correlation length is plotted as a dashed line [see Eq. (\ref{eq_lc_s})] in the left panel. The model for the density correlation length is plotted as a solid line [see. Eq. (\ref{eq_lc_rho})] in the right panel. The measurements from the 512\(^3\) runs are shown in green, those from the 1024\(^3\) runs in blue and those from the 2048\(^3\) runs in red. The measurements from the 1024\(^3\) and \(2048^3\) with turbulence injected at \(L_{\rm box}/14\) are shown in purple and brown respectively. The error bars correspond to the $\pm 1\sigma$ estimated from the time variations of the measured quantities.}
    \label{fig_lc_rho_sigma_Mach_convergence}
 \end{figure*}

\section{Usual estimates of the correlation length}
\label{App_correlation_length}

To estimate the correlation length, several other methods have been suggested in the literature. The first one consists in considering the integral length \citep{Batchelor_TheoryHomogeneousTurbulence1953a} as a proxy of the correlation length. The integral length is defined as
\begin{equation}
    \label{eq_li_Corr}
    l_{\rm i}^\rho = \frac{1}{C_\rho(0)}\int_0^\infty C_\rho(q)dq.
\end{equation}
As justified by Ref. \cite{Jaupart_GeneralizedTransportEquation2021a}, for appriopriate fonctionnal forms of correlation fonctions, \(l_{\rm i}^\rho\simeq l_{\rm c}^\rho\). However, this estimate is only relevant for isotropic turbulence because it is based on the azimuthally averaged ACF \(C_\rho(q)\).

To avoid the computation of the ACF of the density field, which can be numerically expensive in high resolution simulations, we can also use the following estimate of the integral length \citep{Bialy_DrivingScaleDensity2020,Jaupart_GeneralizedTransportEquation2021a}:
\begin{equation}
    \label{eq_li_Var}
    l_{\rm i}^\rho = \frac{L_{\rm box}}{4}\frac{\text{Var}\left(\frac{\Sigma}{\bar{\Sigma}}\right)}{\text{Var}\left(\frac{\rho}{\mathbb{E}(\rho)}\right)},
\end{equation}
where \(\Sigma\) is the column density field along a given line of sight and \(\bar{\Sigma}\) is its spatially averaged value. In a sub-box of size \(L_{\rm box}/2\) with Cartesian coordinates, it is defined by
\begin{equation}
    \Sigma(x, y) = \int_0^{\frac{L_{\rm box}}{2}} \rho(x, y, z) {\rm d}z.
\end{equation}
Again, this estimate is only relevant for isotropic turbulence. For anisotropic turbulence, the result will depend on the choice of the line of sight. We have checked that the two estimates given by Eqs. (\ref{eq_li_Corr}) and (\ref{eq_li_Var}) give the same result.

However, the ratio between the correlation length and the integral length can be as large as 2, which means that the quantity \(M_{\rm inv}\) can be over- or underestimated by almost an order of magnitude. Furthermore, the ratio \(l_{\rm c}^\rho / l_{\rm i}^\rho\) can vary with the Mach number. To calculate \(M_{\rm inv}\) with the quantity \(l_{\rm i}^\rho\) would lead to an apparent violation of the time invariance of this quantity. The study of the properties of \(M_{\rm inv}\) calculated from the integral length is therefore useless.

The consistency of the estimate of the correlation length based on the integral of the ACF until its first zero can also be checked by comparing it with another estimate of the correlation length based on ergodic theory.
\cite{Jaupart_StatisticalPropertiesCorrelation2022} show that the correlation length can be estimated by:
\begin{equation}
    \label{eq_lc_ergo}
    l_c^\rho = \frac{1}{2} \frac{L_{\rm box}}{2}\left(\frac{\mathrm{Var}(\bar{\rho}(t))}{\mathrm{Var}(\rho)}\right)^{1/3}
\end{equation}
where \(\mathrm{Var}(\bar{\rho}(t))\) is the variance of the temporal variation of the volumic estimate of the mean density over time and \(\mathrm{Var}(\rho)\) is the statistical variance of the density field.

This ergodic estimate of the correlation length can only be used for stationary systems. It cannot therefore be used to check the invariance of \(M_{\rm inv}\) in decaying turbulent flows. Furthermore, the convergence of the variance of the statistical variation of the mean density within the box is difficult to achieve. It is necessary to integrate the simulation over a very long time (more than several hundred turbulent crossing times) to have an accurate estimate of this quantity, which is not feasible for the high resolution simulations used in this study.

We also made a comparison in our 1024\(^3\) simulations between the correlation length of the density field computed from the integration of the ACF to its first zeros and that computed from the ergodic theory (Eq. \ref{eq_lc_ergo}) for Mach 3.5 and Mach 10. To obtain a converged result we had to run the simulations (with turbulent forcing) during 70 crossing times. At Mach 3.5 we get \(l_{\rm c}^{1^{\rm st} 0 }/L_{\rm inj}=0.168\pm 0.027\) and \(l_{\rm c}^{\rm ergo}/L_{\rm inj}=0.166\). At Mach 10 we get \(l_{\rm c}^{1^{\rm st} 0 }/L_{\rm inj}=0.126\pm 0.020\) and \(l_{\rm c}^{\rm ergo}/L_{\rm inj}=0.130\). Note that we have no easy way of estimating the uncertainties on \(l_{\rm c}^{\rm ergo}\). These estimates are in very good agreement with each other, confirming the robustness of our calculation of the correlation length using the first zeros of the ACF.

\bibliography{Bib_invariant}

\end{document}